\documentclass[3p,twocolumn]{elsarticle}

\usepackage{graphicx}

\usepackage{amsfonts}

\usepackage{amssymb}
\usepackage{amsmath}
\usepackage{mathtools}
\usepackage{bm}
\usepackage{gensymb}

\usepackage{enumitem}

\usepackage{float}

\usepackage{multicol}

\usepackage[title]{appendix}

\usepackage{lineno}

\usepackage[colorlinks=true,urlcolor=blue]{hyperref} 

\usepackage{dutchcal}

\journal{Energy}

\usepackage{numcompress}\biboptions{authoryear}

\usepackage{nomencl} 

\makenomenclature
\renewcommand*\nompreamble{\begin{multicols}{2}}
	\renewcommand*\nompostamble{\end{multicols}}
\usepackage{framed} 
\usepackage{etoolbox} 
\renewcommand\nomgroup[1]{%
	\item[\bfseries
	\ifstrequal{#1}{A}{Letter symbols}{%
		\ifstrequal{#1}{B}{Greek symbols}{%
			\ifstrequal{#1}{C}{Subscripts and superscripts}{}}}%
	]}

\usepackage{etoolbox}

\renewcommand\nomgroup[1]{%
  \item[\bfseries
  \ifstrequal{#1}{L}{Letter symbols}{%
  \ifstrequal{#1}{G}{Greek symbols}{%
  \ifstrequal{#1}{N}{Non-dimensional numbers}{%
  \ifstrequal{#1}{A}{Abbreviations}{%
  \ifstrequal{#1}{S}{Subscripts and superscripts}{}}}}}%
]}

\begin{document}

\begin{frontmatter}

\title{Real-time data assimilation for the thermodynamic modeling\\of cryogenic storage tanks}

\author[mymainaddress,mysecondaryaddress]{Pedro A. Marques \corref{mycorrespondingauthor}}\cortext[mycorrespondingauthor]{Corresponding author}
\ead{pedro.marques@vki.ac.be}
\author[mymainaddress]{Samuel Ahizi}
\author[mymainaddress]{Miguel A. Mendez}

\address[mymainaddress]{von Karman Institute, Waterloosesteenweg 72, 1640 Sint-Genesius-Rode, Belgium}
\address[mysecondaryaddress]{Université Libre de Bruxelles, Av. Franklin Roosevelt 50, 1050 Bruxelles, Belgium}

\begin{abstract}
The thermal management of cryogenic storage tanks requires advanced control strategies to minimize the boil-off losses produced by heat leakages and sloshing-enhanced heat and mass transfer. This work presents a data-assimilation approach to calibrate a 0D thermodynamic model for cryogenic fuel tanks from data collected in real time from multiple tanks. The model combines energy and mass balance between three control volumes (the ullage vapor, the liquid, and the solid tank) with an Artificial Neural Network (ANN) for predicting the heat transfer coefficients from the current tank state.

The proposed approach combines ideas from traditional data assimilation and multi-environment reinforcement learning, where an agent's training (model assimilation) is carried out simultaneously on multiple environments (systems). The real-time assimilation uses a mini-batch version of the Limited-memory Broyden–Fletcher–Goldfarb–Shanno with bounds (L-BFGS-B) and adjoint-based gradient computation for solving the underlying optimization problem. The approach is tested on synthetic datasets simulating multiple tanks undergoing different operation phases (pressurization, hold, long-term storage, and sloshing). 
The results show that the assimilation is robust against measurement noise and uses it to explore the parameter space further. Moreover, we show that sampling from multiple environments simultaneously accelerates the assimilation. 
\end{abstract}


\end{frontmatter}

\begin{table*}[htb!]  
\begin{framed}

	    \nomenclature[L$A$]{$A$}{area, ${\rm m^2}$}
        \nomenclature[L$A_0$]{$A_0$}{forcing amplitude, ${\rm m}$}
        \nomenclature[L$b$]{$b$}{maximum wave amplitude, ${\rm m}$}
        \nomenclature[L$Cp$]{$C_p$}{isobaric specific heat, ${\rm J/(kgK)}$}
        \nomenclature[L$Cv$]{$C_v$}{isochoric specific heat, ${\rm J/(kgK)}$}
        \nomenclature[L$f$]{$f_e$}{excitation frequency, ${\rm Hz}$}
        \nomenclature[L$f$]{$f_{11}$}{natural frequency, ${\rm Hz}$}
        \nomenclature[L$g$]{$g$}{gravitational acceleration, ${\rm m/s^2}$}
        \nomenclature[L$H$]{$H$}{height, ${\rm m}$}
        \nomenclature[L$h$]{$h$}{heat transfer coefficient, ${\rm W/(m^2 K)}$}
        \nomenclature[L$h$]{$\mathcal{h}$}{mass-specific enthalpy, ${\rm J/kg}$}
        \nomenclature[L$J$]{$\mathcal{J}$}{cost function}
        \nomenclature[L$k$]{$k$}{thermal conductivity, ${\rm W/(mK)}$}
        \nomenclature[L$L$]{$\mathcal{L}$}{Lagrangian function}
        \nomenclature[L$A$]{$\mathcal{A}$}{augmented Lagrangian function}
        \nomenclature[L$L_v$]{$\mathcal{L}_v$}{latent heat of vaporization, ${\rm J/kg}$}
        \nomenclature[L$m$]{$m$}{mass, ${\rm kg}$}
        \nomenclature[L$Q$]{$\dot{Q}$}{heat transfer rate, ${\rm W}$}
        \nomenclature[L$R$]{$R$}{radius, ${\rm m}$}
        \nomenclature[L$T$]{$T$}{temperature, ${\rm K}$}
        \nomenclature[L$t$]{$t$}{time, ${\rm s}$}
        \nomenclature[L$U$]{$U$}{internal energy, ${\rm J}$}
        \nomenclature[L$u$]{$u$}{mass-specific internal energy, ${\rm J/kg}$}
        \nomenclature[L$V$]{$V$}{volume, ${\rm m^3}$}
        \nomenclature[L$W$]{$\dot{W}$}{work, ${\rm W}$}
        \nomenclature[L$p$]{$p$}{pressure, ${\rm Pa}$}
        \nomenclature[L$w$]{$\bm{w}$}{closure parameters}
        \nomenclature[L$R$]{$\mathcal{R}$}{specific ideal gas constant, ${\rm J/kgK}$}
        \nomenclature[L$f$]{$\bm{f}$}{forward model function}
        \nomenclature[L$g$]{$\bm{g}$}{closure law function}
        \nomenclature[L$V$]{$V_m$}{molar volume, ${\rm m^3/mol}$}
        \nomenclature[L$c$]{$\bm{c}$}{radial basis functions coefficients}
        \nomenclature[L$h$]{$\bm{h}$}{observation operator}
        \nomenclature[L$e$]{$\bm{e}$}{observation error}
        \nomenclature[L$R$]{$\bm{R}$}{covariance matrix}
        \nomenclature[L$B$]{$\bm{B}$}{approximation of the inverse Hessian}
        \nomenclature[L$sy_a$]{$\tilde{\bm{s}},\tilde{\bm{y}}$}{reference states and observations}
        \nomenclature[L$sy_b$]{${\bm{s}},{\bm{y}}$}{predicted states and observations}
    
        \nomenclature[G$01$]{$\alpha$}{thermal diffusivity, ${\rm m^2/s}$}
        \nomenclature[G$02$]{$\beta$}{volumetric thermal expansion, ${\rm K^{-1}}$}
        \nomenclature[G$03$]{$\gamma$}{state equation surrogate model}
        \nomenclature[G$04$]{$\delta$}{Peng-Robinson coefficient}
        \nomenclature[G$05$]{$\epsilon$}{width of radial basis functions}
        \nomenclature[G$08$]{$\bm{\theta}$}{vector of parameters}
        \nomenclature[G$08$]{$\kappa$}{kernel of radial basis functions}
        \nomenclature[G$11$]{$\bm{\lambda}$}{adjoint variable}
        \nomenclature[G$13$]{$\nu$}{kinematic viscosity, ${\rm m^2/s}$}
        \nomenclature[G$16$]{$\rho$}{mass-specific density, ${\rm kg/m^3}$}
        \nomenclature[G$17$]{$\sigma$}{ANN activation function}

        \nomenclature[S$v$]{v}{vapor}
        \nomenclature[S$l$]{l}{liquid}
        \nomenclature[S$i$]{i}{interface}
        \nomenclature[S$w$]{w}{walls}
        \nomenclature[S$c$]{c}{critical conditions}
        \nomenclature[S$a$]{a}{ambient}
        \nomenclature[S$sat$]{sat}{saturation}
        \nomenclature[S$ph$]{ph}{phase change}
        \nomenclature[S$k$]{$(k)$}{iteration counter}
        \nomenclature[S$j$]{$\{j\}$}{environment counter}

        \nomenclature[N$Re$]{Re}{Reynolds}
        \nomenclature[N$Pr$]{Pr}{Prandtl}
        \nomenclature[N$Ra$]{Ra}{Rayleigh}
        \nomenclature[N$Gr$]{Gr}{Grashof}
        
	\printnomenclature

\end{framed}
\end{table*}


\section{Introduction}

The current energy crisis has accelerated the interest in sustainable energy sources and cryogenic propellants, liquid hydrogen (LH$_2$) or liquefied natural gas (LNG). These fuels are stored at extremely low temperatures (typically $\approx$-170 $^\circ$C for LNG and $\approx$-250 $^\circ$C LH$_2$). LH$_2$ has been mainly used in rocket engines \cite{Abramson1966} but is now actively explored as an alternative to fossil fuels in many applications, including naval \cite{Grotle2018} and aeronautical \cite{Ball2009} industries. Storage at cryogenic temperatures allows for maximizing the volumetric energy density without resorting to extreme operating pressures ($>300$ bars for gaseous H$_2$)\cite{Fortescue2003, Joseph2016}. The higher thrust-to-weight ratio compared to classical propulsive solutions \cite{Fortescue2003} and the absence of pollutant emission make LH$_2$ a promising energy carrier for a carbon-neutral future \cite{Janic2008, Ball2009}. While the large-scale deployment of LH$_2$ still requires significant technological advancements and infrastructure development, LNG could serve as a cleaner alternative to traditional fossil fuels \cite{LNG} during the transition to a fully renewable energy economy \cite{Ball2009}. 

Nevertheless, storage at cryogenic temperatures requires a complex thermal management system, which poses significant challenges to applications requiring long holding times, from marine to aeronautical, from ground transportation to deep space exploration. No insulating system can entirely prevent heat exchanges with the surroundings; thus, some liquid unavoidably evaporates over time and increases the tank pressure \cite{Petitpas2018, duan_thermal_2021, perez_2021}. 
Additional challenges are faced in tanks installed on vehicles, as external accelerations induce sloshing. Sloshing, defined as the movement of the free liquid surface, can increase heat and mass transfer rates between the liquid and ullage gasses and thus produce significant variations of the tank’s pressure \cite{Arndt2011, Ludwig2013}.

The heat and mass transfer exchanges in cryogenic tanks pose a challenging task in developing accurate modeling tools for these systems. Numerous studies \cite{Petitpas2018, Osipov2011, Migliore2016, Grotle2018, Wang2020, Jo2021, Marques2022} resort to dynamical models derived from conservation laws in a quasi-dimensional (0D) framework. The tank is described in these models as a set of multiple control volumes (CVs) that can exchange energy and mass. These models are straightforward to derive and computationally light, hence easy to integrate with other sub-models to perform real-time system-wide simulations for fault detection \cite{Daigle2011}, process optimization \cite{Jiang2021, Wang2021, Tian2021}, or control applications \cite{Kalikatzarakis2022}.

All simplified models depend on closure laws to predict boil-off rates, heat losses to the environment, and heat transfer within the system. To this end, most authors employ empirical correlations to express these processes in stationary and long-term storage conditions. However, correlations accounting for the wide range of sloshing, refilling, pressurization, controlled venting, and all possible scenarios encountered by a cryogenic tank do not exist. Focusing on the case of sloshing, Ludwig \& Dreyer \cite{Ludwig2013} derived correlations for heat transfer linking a sloshing-based Nusselt number to sloshing amplitude and frequency of the motion, but these correlations are only valid for a limited range of sloshing regimes. A general inverse model-based method to extract 
heat transfer correlations from real-time measurements of pressure, temperature, and liquid level in a tank undergoing sloshing was developed by the authors in \cite{Marques2022}. This inverse modeling approach can be seen as an example of data assimilation \cite{Carrassi2017,cheng_machine_2023}, a framework for optimally integrating data and numerical models. In the context of this work, the assimilation consists of identifying heat and mass transfer laws such that the prediction of a 0D thermodynamic model matches with real-time data as closely as possible.


This work expands the approach in \cite{Marques2022} combining assimilation and machine learning to bring two main novelties.
First, the optimization underlying the assimilation is solved using a combination of adjoint method \cite{adjoint} and a mini-batch version of the L-BFGS-B (Limited-memory Broyden–Fletcher–Goldfarb–Shanno with bounds) algorithm \cite{LBFGSB}. Second, perhaps more interestingly, we combine we introduce a multi-environment reinforcement learning formalism, to carry out the assimilation from multiple tank simultaneously even if these undergo independent thermodynamic evolution and operate in widely different conditions.

The proposed approach takes inspiration from multi-agent and multi-environment reinforcement learning (MARL/MERL) \cite{Zhang2021}. The reader is referred to \cite{ifaei_sustainable_2023} for an extensive overview of machine learning algorithms and their application to the development of sustainable energies.

In reinforcement learning \cite{sutton2018reinforcement,Bertsekas2019,Pino2023}, an \emph{agent} (e.g. robots, autonomous systems or software applications) learns to make sequential decisions in an \emph{environment} to achieve a goal. The agent is usually an Artificial Neural Network (ANN \cite{goodfellow2016deep}) that maps states to actions. For example, in developing a software application for playing chess, the agent's input could be images of the chessboard status, and the agent's output (action) is the next best move. In multi-environment learning, the agent learns by interacting with many environments (e.g. playing multiple chess games in parallel). In multi-agent learning, agents interact with each other and the environment to achieve a common goal (not applicable to the chess example, but relevant to team games). This is an active area of research with many open questions and challenges \cite{Buşoniu2010,ifaei_sustainable_2023} on handling communication and coordination between agents and learning effective policies in large-scale multi-agent systems.

We believe that the combination of modern variational assimilation tools with the multi-environment formalism in reinforcement learning has enormous potential in the modeling, optimization and control of energy systems. This work moves the first step towards its application to the problem of data-driven calibration of thermodynamic models for cryogenic tanks. In this context, the agent's input is real-time measurements of the tank's state (e.g., pressure, temperature, fill level, etc), and the action is the prediction of the heat transfer coefficients, which are then passed to a 0D model that predicts the following states. The comparison between prediction and state provides feedback to the agent. The multi-environment formalism allows the agent to interact (and learn from) multiple tanks, much like a machine learning agent learns to play chess playing multiple games simultaneously.

In this work we present the mathematical formalism and provide a first proof of concept using "virtual environments" for which the ground truth is available. We focus on the feasibility of the learning process, its convergence and sensitivity to measurement noise and we study how the use of multiple environments favors the learning in various scenarios (tank pressurization, hold, long term storage and sloshing). 

This article is structured as follows. Section \ref{sec2} outlines the physical modeling of the cryogenic fuel tank implemented in this work. This was complemented with a surrogate model for the thermodynamic properties, described in section \ref{sec3}. Section \ref{sec4} describes the key physical phenomena tackled in this work and details how they are modeled in the current framework. The real-time data assimilation and inverse method strategies are outlined in Section \ref{sec5}. Furthermore, this section also describes how the synthetic test cases were generated.
Lastly, Section \ref{sec6} overviews the results of the model calibration in single-environment and multi-environment conditions. Concluding remarks and future outlook are discussed in Section \ref{sec7}.

\section{Thermodynamic modeling}
\label{sec2}

\begin{figure}[h]
	\centering
	\includegraphics[width=0.85\linewidth]{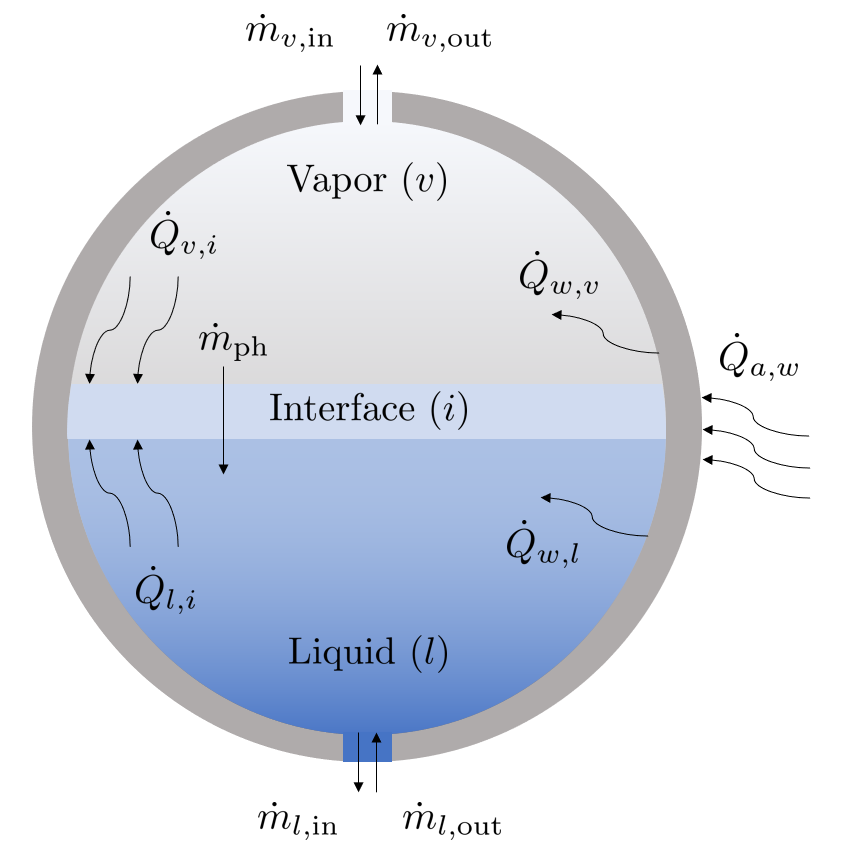}
	\caption{Schematic of the cryogenic fuel tank and its subsystems: vapor $(v)$, liquid $(l)$, and insulating walls $(w)$. The heat and mass exchanges between these control volumes are expressed through the $\dot{m}$ and $\dot{Q}$ fluxes.}
	\label{fig:0D_schematic}
\end{figure}

The cryogenic storage tank considered in this work is a single-species system composed of a liquid and its vapor enclosed in  insulated walls. Figure \ref{fig:0D_schematic} provides a schematic of the problem with the relevant parameters involved. The subscripts $l$, $v$, and $w$ are used to distinguish variables related to the liquid, the vapor, and the wall, respectively. 

The gas-liquid interface separating vapor and liquid is treated as an infinitesimally thin region where heat and mass transfer occurs. The system exchanges heat and mass, both on the liquid and the vapor side, during various operations (e.g., pressurization, venting, filling). The reader is referred to the list of symbols for the nomenclature.

The 0D thermodynamic model used in the assimilation expresses the conservation of mass and energy in three control volumes: the liquid, the vapor and the solid walls. Considering mass-averaged thermodynamic properties, these balances result in a system of ordinary differential equations (ODEs), which must be closed with empirical relations for the heat and mass transfer rates. Finding closure from real-time data is the objective of the data assimilation approach proposed in this work. 

The mass conservation between the liquid and the vapor is given by 

\begin{align}
    \label{eq:dmvdt}
    \frac{dm_v}{dt} &= \sum \dot{m}_{v,\text{in}} - \sum \dot{m}_{v,\text{out}} - \dot{m}_\text{ph}\\
    \frac{dm_l}{dt} &= \sum \dot{m}_{l,\text{in}} - \sum \dot{m}_{l,\text{out}} + \dot{m}_\text{ph}
\end{align}

\noindent
where $\dot{m}_\text{ph}=\dot{m}_\text{cond} - \dot{m}_\text{evap}$ is the net mass flux through the interface, accounting for the balance of condensation and evaporation. The mass conservation in the vapor and liquid phases give

\begin{equation}
\begin{aligned}
\label{eq:duvdt}
    \frac{dU_v}{dt} =& 
    \sum \dot{m}_{v,\text{in}}\mathcal{h}_{v,\text{in}}    - 
    \sum \dot{m}_{v,\text{out}}\mathcal{h}_{v,\text{out}}\\
    &-\dot{m}_\text{ph}\mathcal{h}_{v,\text{sat}}
     -\dot{Q}_{v,i} + \dot{Q}_{w,v} + \dot{W}_v\,, \\[1em]
\end{aligned}
\end{equation}

and 

\begin{equation}
\begin{aligned}
\label{eq:duldt}
    \frac{dU_l}{dt} =& 
    \sum \dot{m}_{l,\text{in}}\mathcal{h}_{l,\text{in}}    -
    \sum \dot{m}_{l,\text{out}}\mathcal{h}_{l,\text{out}}\\
    &+\dot{m}_\text{ph}\mathcal{h}_{v,\text{sat}}
     -\dot{Q}_{l,i} + \dot{Q}_{w,l} + \dot{W}_l\,,
\end{aligned}
\end{equation}where $U_v$ and $U_l$ are the internal energies of the vapor and liquid phases, $\dot{Q}$ denotes the general heat flux, $\mathcal{h}$ is the specific enthalpy, and $\dot{W}=-pdV/dt$ is the expansion/compression work due to changes in filling level.

The model closure is required to link the heat transfer rates at the interface. Assuming that the heat transfer occurs at much larger time scales than the interface dynamics, we consider quasi-steady formulation and use Newton's cooling law for the closure relation:

\begin{equation}
\begin{aligned}
\label{eq:Q}
    \dot{Q}_{v,i} &= A_i h_{v,i} (T_v - T_i) 
    \\
    \dot{Q}_{w,v} &= A_{w,v} h_{w,v} (T_w - T_v) 
    \\
    \dot{Q}_{l,i} &= A_i h_{l,i} (T_l - T_i)
    \\ 
    \dot{Q}_{w,l}  &= A_{w,l} h_{w,l} (T_w - T_l),
\end{aligned}
\end{equation}

\noindent
where $h_{v,i}$, $h_{l,i}$, $h_{w,v}$, $h_{w,l}$ are the heat transfer coefficients, $A_i$ is the gas-liquid interface area, $A_{w,v}$ and $A_{w,l}$ are the surface exchange areas between the walls-vapor and walls-liquid phases. 
The heat transfer coefficients are unknown and must be identified from the data. Defining these model parameters as $\bm{\theta}=[h_{v,i},h_{l,i},h_{w,v},h_{w,l}]$, we assume that a closure parametric relation can be used to link to the state of the thermodynamic model.

The interface temperature in \eqref{eq:Q} is assumed to be the saturation temperature evaluated at vapor pressure $p_v$; hence the energy balance at the interface provides the mass flux due to phase change as

\begin{equation}
    \dot{m}_\text{ph} = \frac{\dot{Q}_{l,i} - \dot{Q}_{v,i}}{\mathcal{L}_v}\,,
\end{equation}

\noindent
with $\mathcal{L}_v$ the latent heat of vaporization. 

Concerning the exchange areas in \eqref{eq:Q}, we do not account for the time variation of $A_i$ (due to, e.g., sloshing), and take it as the tank's cross-section when the tank is half-filled. On the other hand, given the tank's geometry, the areas $A_{w,v}$ and $A_{w,l}$ are updated at each time step depending on the liquid level. This can be computed from the liquid and vapor masses and their properties. In particular, treating the liquid phase as incompressible (as in \cite{vanForeest2014}), its density is solely a function of temperature $\rho_l (p,T)\approx \rho(T)$, hence

\begin{equation}
    \frac{dV_l}{dt} = - \frac{dV_v}{dt} \approx
    \frac{1}{\rho_l} \frac{dm_l}{dt}-
    \frac{m_l}{c_{p,l}\rho_l^2}
    \left(\frac{d\rho_l}{dT}\right)_p\frac{du_l}{dt},
\end{equation} 

\noindent

where $c_{p}$ is the specific heat at constant pressure, $\rho$ is the density, and $u=U/m$ is the specific internal energy. Finally, in the 0D formulation, the insulating walls are treated as a single control volume which can exchange heat with the vapor and liquid phases, as well as with the external environment at ambient conditions. For a single control volume with mass-averaged properties, the internal energy of the walls evolve as

\begin{equation}
    \label{eq:duwdt}
    \frac{dU_w}{dt} = \dot{Q}_{a,w} - \dot{Q}_{w,v} - \dot{Q}_{w,l}
\end{equation}

\noindent
where $\dot{Q}_{a,w}$ is the heat entering the tank from the environment. This term could be computed from empirical correlations \cite{Incropera2007,huerta_realistic_2019, Jo2021}, but in this work, this is taken as a user-defined function that depends on the specific scenario experienced by the tank (described in the following section). It is worth noticing that equation \eqref{eq:duwdt} can be written as a function of the mass-averaged solid temperature by introducing $dU_w=m_{w}c_wdT_w$.

All thermodynamic properties ($c_p$, $\rho$) and the link between internal energies, temperatures and pressures are
evaluated using the surrogate model described in Section \ref{sec3}.


To conclude this section, we note that the thermodynamic model is constituted of equations \eqref{eq:dmvdt}-\eqref{eq:duwdt} and can be cast in the form of a parametric initial value problem:

\begin{equation}
\label{eq:dxdt}
\left\{ \begin{aligned} 
  \frac{d\bm{s}}{dt} &= \bm{f}\left(\bm{s},t;\bm{\theta}\right) \\
  \bm{\theta}&=\bm{g}(\bm{s};\bm{w})\\
  \bm{s}(0) &= \bm{s}_0
\end{aligned} \right.
\end{equation}

\noindent
where $\bm{s} = \left[ m_v, m_l, u_v, u_l, T_w, V_v, V_l \right] \in\mathbb{R}^{7}$ is the state vector (including the masses, volumes and internal energies of each control volume), describing the thermodynamic condition of the system at time-instant $t$, $\bm{s}_0\in\mathbb{R}^{7}$ is the (known) initial condition,  
$\bm{\theta}\in\mathbb{R}^{4}$ is the vector of model parameters (heat transfer coefficients in eq. \ref{eq:Q}), and $\bm{g}:\mathbb{R}^{7}\rightarrow \mathbb{R}^4$ is the closure law that depends on a set of closure parameters $\bm{w}\in\mathbb{R}^{n_w}$. Finding these parameters from on-line data is the essence of the data assimilation problem investigated in this work. 

We use empirical correlations introduced in Section \ref{sec5p1} to generate synthetic data and an artificial neural network Section \ref{sec5p2} as a general purpose parametric function approximator for solving the assimilation in real-time.


\section{Surrogate of the real fluid properties}\label{sec3}

A reliable tool to compute the real fluid properties as well as all the relevant derivatives involved in the thermodynamic modeling is the CoolProp package \cite{CoolProp}. This library implements pure fluid equations of state and transport properties using the Helmholtz energy formulations. However, the data assimilation approach used in this work requires a large number of calls to this package, resulting in a considerable computational cost.

To reduce this cost, we propose a surrogate model of the equation of states. This model is formulated as a correction of the Peng-Robinson model \cite{Peng1976} combined with a first order Taylor expansion for the internal energy. Specifically, given

\begin{equation}
    p^* = \frac{\mathcal{R}T}{V_m - b} - \frac{a\delta}{V_m^2 + 2bV_m - b^2}
\end{equation} the pressure computed from Peng-Robinson's model, with $v=1/\rho$ the specific volume, $V_m$ the molar volume, $a=0.45724{R^2T_c^2}/{p_c}$, $b=0.07780{\mathcal{R} T_c}/{p_c}$ and $\delta = 1.202e^{-0.30288T/T_c}$, the first order expansion for the temperature, centered at ($p_\text{ref},T_\text{ref}$) gives

\begin{equation}
\label{Taylor}
   \Delta T^*  = \frac{1}{c_{v_\text{ref}}} \Biggl \{{\Delta u - \Bigl[T\biggr(\frac{\partial p}{\partial T}\biggl)_v - p_\text{ref}\Bigr]\Delta v}\Biggr\}\,,
\end{equation} such that $T^*=T_\text{ref}+\Delta T^*$.

The surrogate model for predicting the fluid properties in the thermodynamic model are then written as 

\begin{equation}
\begin{aligned}
     p = p^*+\gamma_p(u,\rho), \,
     T = T^* +\gamma_T(u,\rho)\,,\\
     T_\text{sat} = \gamma_{S}(u,\rho)\,,\\
     \mathcal{h}_{g,\text{sat}}=\gamma_{\mathcal{h},g}(u,\rho)\,,
     \mathcal{h}_{v,\text{sat}}=\gamma_{h,v}(u,\rho)\,.
\end{aligned}
\end{equation} 

\begin{figure}[h]
	\centering
	\includegraphics[width=0.85\linewidth]{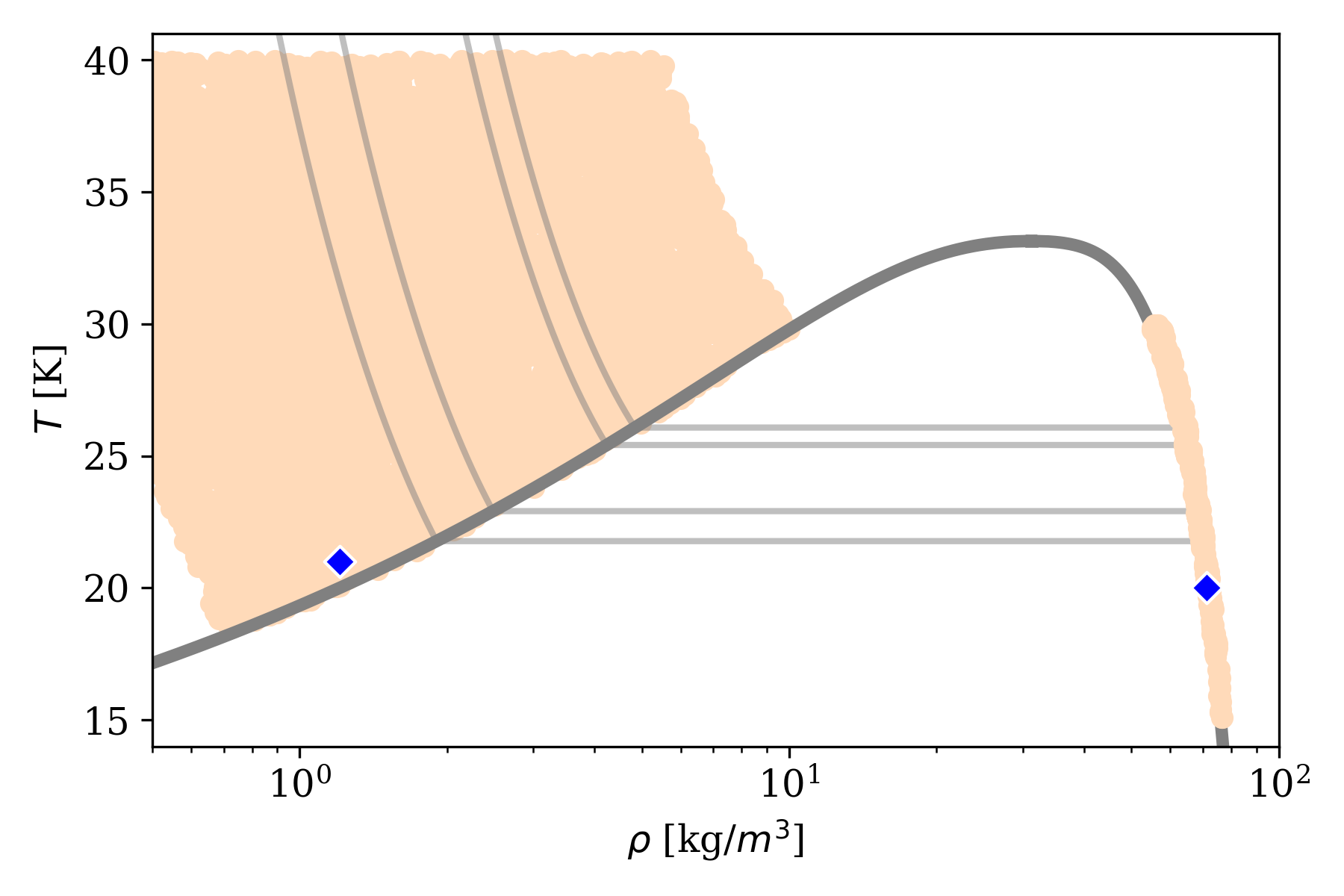}
	\caption{Range of $(T,\rho)$ values spanned by the simulations (in orange) and coordinates of the reference point used in the \eqref{Taylor} for the temperature in the liquid and gaseous phases (blue diamonds).}
	\label{REF_RHO_P}
\end{figure}

The surrogate models $\gamma_p,\gamma_T,\gamma_S,\gamma_{h,g},\gamma_{\mathcal{h},v}$ are constructed using Radial Basis Function (RBF) interpolation from a dataset of 500 points built from Coolprop. These are randomly placed in the area of the $(u,\rho)$ plane spanned by the simulations discussed in section \ref{sec6}. These consider the storing of LH$_2$ and the relevant region is shown in Figure \ref{REF_RHO_P}.
The reference point for the Taylor expansion in \eqref{Taylor} is shown with a blue diamond.

To build the regression, all the input points in the space $(u,\rho)$ are min-max transformed in the range $[0,1]$ (recall that the min-max transform of the variable $x_i$ is $\tilde{x_i} = (x_i - \text{min}(x_i) / (\text{max}(x_i) - \text{min}(x_i)))$) while the outputs are log-transformed as $\Tilde{y_i} = \ln(|y_i| + 1)$ prior to the fitting of the RBF interpolator.

The interpolator in the scaled $\mathbb{R}^2$ domain is built using Gaussian RBF kernel $\kappa(r) = e^{-r^2/\varepsilon^2}$, with $r=||(\tilde{\rho},\tilde{u})-(\tilde{\rho}_i,\tilde{u}_i)||_2$ the $l_2$ norm of the distance between points $(\tilde{\rho},\tilde{u})$ and training data $(\tilde{\rho}_i,\tilde{u}_i)$ and $\varepsilon=0.01$. To describe the training/prediction step of the RBF interpolator, we define as $\tilde{X},\tilde{Y}$ the set of $n_*$ training points and let ${Y}\in\mathbb{R}^{n_{**} \times 6}$ be the model prediction for any of the interpolated variables in unseen points ${X}\in\mathbb{R}^{n_{**}\times 2}$.

Given $\Phi(\tilde{X},{X})=\kappa(\bm{D}(\Tilde{X},{{X}}))$ 
the matrix collecting the RBF kernel evaluated at the set of locations ${X}$, with $\bm{D}(\Tilde{X},{X})$ the matrix of distances between the training points $\Tilde{X}$ and the points ${X}$, the training consists in finding the set of coefficients

\begin{equation}
\bm{c}=\left(\Phi(\Tilde{X},\Tilde{X})^T \Phi(\Tilde{X},\Tilde{X})\right)^{-1}\Phi(\Tilde{X},\Tilde{X})^T \Tilde{Y}
\end{equation} such that the predictions on $X$ can be computed as 

\begin{equation}
Y = \Phi(\Tilde{X},{X})\,\bm{c}\,.
\end{equation}

In this work, the number of training points is $n_{*}=500$ and the predictions can be requested on a set of points  $n_{**}$ simultaneously. This is particularly convenient for the assimilation strategy.
Moreover, the interest in RBF interpolation is that derivatives are analytically available by replacing the kernel function with its derivatives and using the chain rule. We skip the details of this computation in the 
interest of brevity. The surrogate model is approximately four times faster than Coolprop and offers an approximation that differs from the Helmholtz equation's prediction by less than $1\%$ within the investigated domain.

\begin{figure}[h]
	\centering
	\includegraphics[width=1.0\linewidth]{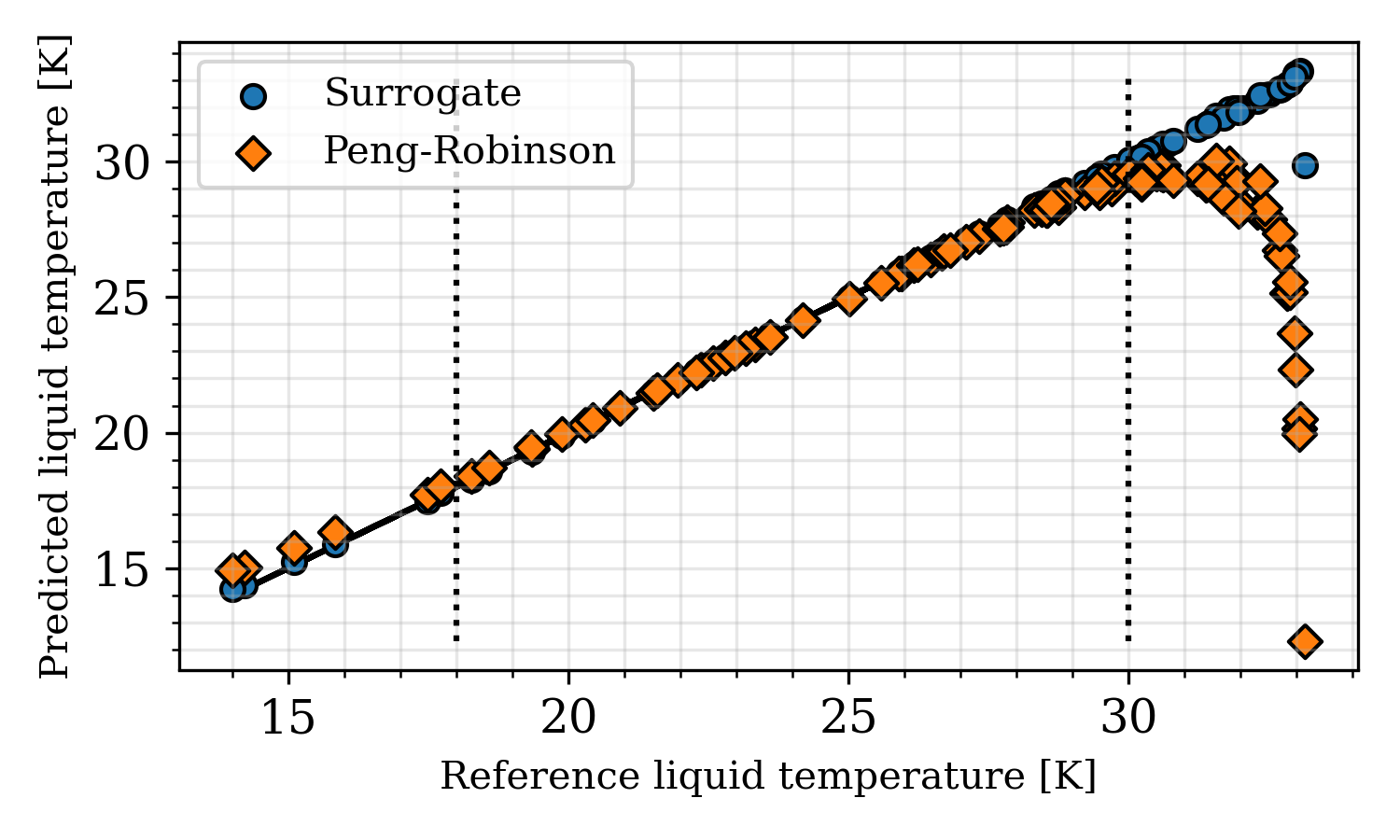}
	\caption{Liquid temperature prediction using Peng-Robinson and the surrogate against reference computed using CoolProp.}
	\label{fig:Tl_surrogate}
\end{figure}

For illustrative purposes, Figure \ref{fig:Tl_surrogate} compares the prediction of the surrogate model (blue circles) with the reference data from Coolprop for the liquid temperature over a broad range of data randomly sampled within the area of interest in Figure \ref{REF_RHO_P}. The predictions are compared with the ones from Peng-Robinson's model. Contrary to Peng-Robinson's model, the surrogate model remains accurate even close to the critical conditions.

\section{Investigated thermodynamic scenarios}
\label{sec4}

Cryogenic fuel tanks have standard handling sequences consisting of tank chilling, filling, boil-off, level adjustment, pressurization, and hold \cite{Joseph2016}. The scenarios considered are described below.

\textbf{Tank pressurization.} This is achieved by injecting vapor or inert gas in the ullage of the tank (active pressurization) or by introducing heat (passive pressurization) to produce a controlled fuel boil-off. Active pressurization is faster than passive boil-off \cite{Arndt2011} and is often preferred to avoid warming up the liquid.
In the proposed 0D model, active pressurization is achieved by a positive mass flux $\dot{m}_{v,\text{in}}$ with enthalpy $\mathcal{h}_{v,\text{in}}$, evaluated at a certain pressure and temperature. Passive pressurization is achieved by providing heat $\dot{Q}_{a,w}$.

\textbf{Venting.} This operation is performed to reduce the ullage pressure after a pressure rise event and to limit the self-pressurization due to heat ingress. In absence of dedicated control systems (e.g.the Thermodynamic Venting in \cite{Qin2021,Imai2020}), this is achieve by simply expelling vapor to the outside. In the 0D model we thus model the vengint event with a negative mass flux $\dot{m}_{v,\text{out}}$ with enthalpy $\mathcal{h}_{v,\text{out}}$, evaluated at the ullage conditions.

\textbf{Level adjustment.} These operations are performed by either injecting or removing liquid from the tank. This is modeled by imposing a positive $\dot{m}_{l,\text{in}}$ with enthalpy $\mathcal{h}_{l,\text{in}}$ in case of a refilling (evaluated at a prescribed pressure and temperature), or a negative $\dot{m}_{l,\text{out}}$ with enthalpy $\mathcal{h}_{l,\text{out}}$ for liquid extraction (evaluated at the liquid conditions). 

\textbf{Hold.} This phase refers to the period during which the tank is maintained in static conditions, with no addition or removal of fuel. Once the tank has been pressurized, the hold phase is characterized by a gradual decrease (relaxation) of the pressure. This is partly attributed to condensation and the heat exchanges with the colder interface and inner tank walls \cite{Ludwig2013}. The pressure drop during this phase depends on the fluid properties, the pressurization method, duration, and the tank volume \cite{Arndt2011}. Within the 0D formulation in this work, these exchanges can be modeled by an appropriate definition of $h_{v,i}$, $h_{l,i}$, $h_{w,v}$, $h_{w,l}$.

\textbf{Long-term storage.} This is similar to the previous one but differs in the time scale. In long-term storage, heat leaks in from the surrounding and increases the temperature of the inner walls in contact with the fuel. The resulting buoyancy-driven heat fluxes create a thermal stratification in the gas and liquid phases \cite{Petitpas2018}. The main heat transfer mechanisms inside the tank are (1) wall-normal fluxes entering the liquid, transporting warm fluid upwards, and causing a thermal boundary layer to develop at the interface, (2) wall tangent fluxes due to the vertical temperature gradient along the solid walls, (3) exchanges between the warmer gas and the colder liquid, which can trigger condensation or evaporation and promote thermal stratification \cite{vanForeest2014}. In the 0D model, we mimic the impact of these thermal gradients by adjusting the heat transfer coefficients between the subsystems. In addition (and differently from the hold phase), we include an external heat flux $\dot{Q}_{a,w}$.

\begin{figure*}[h]
	\centering
	\includegraphics[width=0.8\linewidth]{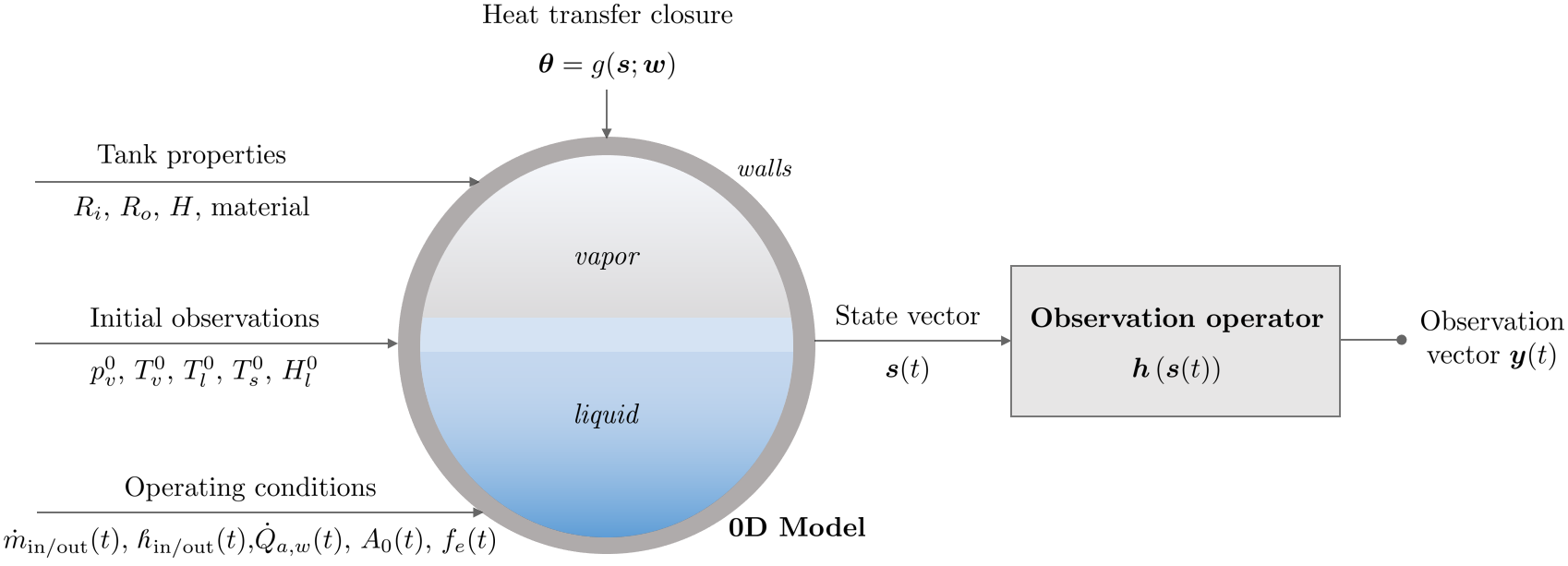}
	\caption{Schematic of the parameters required to define an environment for the data assimilation of the 0D thermodynamic model of the cryogenic tank. Interaction with the environment produces the observation vector $\bm{y}(t)$ simulating measurements from the tank.}
	\label{fig:synthetic_data_generation_approach}
\end{figure*}

\textbf{Sloshing.} External accelerations trigger liquid sloshing in the tank. This results in significant enhancement of heat and mass transfer, which in turn produces large pressure oscillations. The most relevant sloshing-induced thermodynamic phenomenon is the so-called ‘pressure drop effect’\cite{Arndt2011,Ludwig2013,Marques2022,wang_dynamic_2021} occurring when the sloshing reduces the average saturation temperature at the gas-liquid interface and consequently triggers condensation of the ullage vapor. The opposite mechanism can also occur when the walls are well above the saturation temperatures, hence the liquid near the contact line evaporates or boils, producing a pressure rise \cite{Arndt2011}. This complex interaction between fluid dynamics, heat transfer, and phase change poses a significant challenge to current propellant management strategies. In our 0D modeling, sloshing is simulated by modifying the heat transfer coefficients. These depend on the intensity of the dynamic perturbations imposed on the fuel tank.



\section{Real-time data assimilation framework}
\label{sec5}

We describe the virtual environment generating the synthetic database in section \ref{sec5p1} and the multi-environment real-time assimilation in sections \ref{sec5p2} and \ref{sec5p3}.

\subsection{Environment Simulation}
\label{sec5p1}

Borrowing from the reinforcement learning terminology, the assimilation is carried out by an \emph{agent} interacting with an \emph{environment} to achieve a goal. In our context, the \emph{agent} is a function that must predict the closure parameters (heat transfer coefficients) with the goal of having the model prediction as close as possible to the available data. Therefore, the \emph{environment} definition requires defining (1) the geometry of the fuel tanks, (2) the working fluid, (3) the initial observations of the system, (4) the temporal evolution of the heat transfer coefficients (unknown to the agent), and (5) the sequence of operations applied to the tank. Figure \ref{fig:synthetic_data_generation_approach} summarizes the relevant quantities.

The multi-environment formulation involves learning the \emph{same} closure law from multiple environments. In this work, we consider multiple cryogenic fuel tanks, partly filled with LH$_2$. In all environments, the heat transfer coefficients in \eqref{eq:Q} are modeled as follows

\begin{equation}
\begin{aligned}
\label{eq:h_all}
    \theta_0 = h_{v,i} &= \frac{k_v}{R}\left( 140\, \text{Re}_{s,v}^{0.69} \text{Pr}_v^{1/3} + 10\, \text{Ra}_v^{0.15} \right)
    \\
    \theta_1 = h_{l,i} &= \frac{k_l}{R}\left( 50\,\, \text{Re}_{s,l}^{0.69} \text{Pr}_v^{1/3} + 2\, \text{Ra}_l^{0.15} \right)
    \\
    \theta_2 = h_{w,v} &= 20\, h_{v,i}
    \\
    \theta_3 = h_{w,l} &= 10\, h_{l,i}
\end{aligned}
\end{equation}

\noindent
where $\text{Pr}=\nu/\alpha$ is the Prandtl number, $\text{Re}_s$ is the Reynolds number due to sloshing and  $\text{Ra}$ is the Rayleigh number. These are defined as

\begin{align}
\label{eq:DIMs}
    \text{Re}_{s} &= \frac{f_e}{f_{11}}\left(\frac{b}{R}\right)^2\frac{\left( g R^3 \right)^{1/2}}{\nu} \sqrt{1.841}\,, \\ 
    \text{Ra} &= \frac{g\beta\Delta T R^3}{\nu \alpha}\,,
\end{align}

\noindent
where $f_e$ is the frequency of the sinusoidal forcing motion acting on the tank, $f_{11}$ is the natural frequency of the tank \cite{Abramson1966}, $b$ is the maximum expected wave-height during sloshing, which is a function of the forcing amplitude $A_0$ \cite{Ludwig2013}, $R$ is the tank's radius, $\nu$ is the kinematic viscosity, $g$ is the gravitational acceleration, $\beta$ is the volumetric thermal expansion coefficient, and $\alpha$ is thermal diffusivity. 

The $h_{l,i}$ and $h_{v,i}$ coefficients account for forced convection through the Reynolds and Prandtl numbers, and buoyancy-driven fluxes through the Rayleigh number. 
The relations presented in \eqref{eq:h_all} were tuned to roughly portray the experimental pressure and temperature reported in \cite{Arndt2011,Ludwig2014,Moran1994,Hassan1991} and are unknown to the agent.

We denote as $\tilde{\bm{s}}(t)$ the state of the environment from which the assimilation is carried out and as $\bm{s}(t)$ the prediction of the model using the closure law provided by the agent, as described in Section \ref{sec5p2}. We assume that the interaction between agent and environment is carried out by monitoring some observations of the system, here denoted as $\tilde{\bm{y}}(t;\bm{w}) = \bm{h}(\tilde{\bm{s}}(t;\bm{w}))$, with $\bm{h}(\cdot)$ the observation function simulating a measurement process. The observations considered in this work are the vapor pressure $p_v(t)$, the mass-averaged vapor $T_v(t)$, liquid $T_l(t)$, and wall $T_w(t)$ temperatures, and the fill-level $H_l(t)$. These quantities were retrieved from the state vector $\bm{s}$, Hence the observation function is $\bm{h}(\cdot):\mathbb{R}^7\rightarrow\mathbb{R}^5$.

\subsection{Agent definition}
\label{sec5p2}

 We define an agent using a feed-forward Artificial Neural Network (ANN). This architecture comprises a large number of interconnected nodes, called neurons, organized in layers \cite{Goodfellow2016}. The input nodes receive data and propagate it through hidden layers until the output nodes.

The inputs of this parametric model are the Prandtl numbers $\text{Pr}$ of the gaseous and liquid phases, the tank's aspect ratio $R/H$, the Reynolds number defined from the excitation velocity $\text{Re}_l = {f_eA_eR}/{\nu}$ and three Grashof numbers $\text{Gr} = {g\beta\Delta TR^3}/{\nu^2}$ computed with the temperature differences between the three control volumes (solid walls, gas and liquid). The inputs are further transformed: a logarithmic transform is applied to the Grashof number $\widetilde{\text{Gr}}= \ln(|\text{Gr}|+1)$, while the Reynolds number is divided by the limit of the full turbulent transition $\widetilde{\text{Re}}=\text{Re}_l/\text{Re}_c=\text{Re}_l/10^4$. 
Denoting as $\bm{x}$ and $\bm{\theta}$ the vectors of inputs and outputs of the ANN, the general architecture can be defined recursively as

\begin{equation}
    \bm{\theta} = \sigma^{(L)}{( \bm{z}^{(L-1)} )}
\end{equation}

with 

\begin{equation}
    \begin{cases}
			\bm{a}^{(1)}=\bm{x} & \\
            \bm{a}^{(l)}=\sigma^{(l)}\left (\bm{z}^{(l)}\right) & \\
            \bm{z}^{(l)}=\bm{W}^{(l-1)}\bm{a}^{(l-1)}+\bm{b}^{(l)} & 
		 \end{cases}
\end{equation}

\noindent
where $l=1,2,...,L$ is the span of the layers, $\mathbf{a}^{(l)},\mathbf{b}^{(l)}\in\mathbb{R}^{n_l}$ are the activation and the bias vectors per layer, $n_l$ is the number of neurons per layer, $\sigma$ is the activation function , $\mathbf{W}^{(l)}\in\mathbb{R}^{n_l\times n_{l-1}}$ is the matrix containing the weights connecting layer $l-1$ with layer $l$.

The neural network employed in this work consists of $L=2$ layers of 16 neurons ($n_l=8$) with Rectified Linear (ReLu) activation functions, which results to a total of $n_w=484$ weights and biases parameters. These are arranged into the vector of parameters $\bm{w}\in \mathbb{R}^{n_w}$ introduced in \eqref{eq:dxdt}.

The scope of data assimilation consists in training this ANN (i.e. identify the parameters parameters $\bm{w}\in \mathbb{R}^{n_w}$) using real time data collected from multiple environments. The optimal set of parameters is the one that provides the best match on the predictions of the thermodynamic state, hence such that the simulated observations $\bm{y}(t)=\bm{h}(\bm{s}(t))$ match with the true observations $\tilde{\bm{y}}(t)=\bm{h}(\bm{s}(t))$.

\subsection{Multi-environment assimilation}
\label{sec5p3}

 To quantify the performances of the assimilation, we define a cost function $\mathcal{J}(\bm{w})$ measuring the discrepancy between model prediction and data across all the available environments. Denoting as $\bm{y}_{\{j\}}(t)$ the observations predicted bt the agent in the $j^\text{th}$ environment and as $\tilde{\bm{y}}_{\{j\}}(t)$ the collected ones, the cost function is 
 
\begin{equation}
\begin{aligned}
	\label{eq:J}
        \mathcal{J}(\bm{w}) &=
        \frac{1}{N}
        \sum_{j=1}^{N}        
        \int_{t_i}^{t_i+T} \mathcal{L}\left(\tilde{\bm{y}}_{\{j\}},\bm{y}_{\{j\}},t\right) dt \\
        &= 
        \frac{1}{2 N}
        \sum_{j=1}^{N}        
        \int_{t_i}^{t_i+T} 
        \bm{e}_{\{j\}}(t)^T \bm{R}_{\{j\}}^{-1} \bm{e}_{\{j\}}(t) dt       
\end{aligned}
\end{equation}

\noindent
where $N$ is the number of environments, $T$ is the observation time, $\bm{e}_{\{j\}}(t) = \tilde{\bm{y}}_{\{j\}}(t) - \bm{y}(t)$ is the prediction error and $\bm{R}_{\{j\}}$ is the covariance matrix accounting for measurement noise. Since the assimilation seeks to learn the full set of parameters from all environments, the proposed formulation is a multiple-environment but single-agent framework, as opposed to a multi-agent formulation in which different agents could be assigned to learn different coefficients in the same environment.

In this work, the function $ \mathcal{J}(\bm{w})$ is minimized through gradient-based optimization using the L-BFGS-B optimizer \cite{LBFGSB} coupled with a variational (adjoint-based \cite{adjoint}) method to compute the gradient $\nabla_{\bm{w}} \mathcal{J}$. 

The adjoint method allows computing this gradient without computing the sensitivity of the state with respect to the closure parameters, i.e. $d \bm{s}/d\bm{w}$. The gradient is computed by relying on the augmented Lagrangian function $\mathcal{A}(\bm{w})$

\begin{equation}
\begin{aligned}
\label{eq:L}
    \mathcal{A}(\bm{w})
    &=\\
    &\sum_{j=1}^{N}
    \Biggl(
        \int_{t_i}^{t_i+T}
        \mathcal{L} (\bm{w}) + \\
        &
        \bm{\lambda}_{\{j\}}(t)^T\Biggl(
        \bm{f}(\bm{s}_{\{j\}},t;\bm{w}) -
        \frac{d\bm{s}_{\{j\}}}{dt}
        \Biggr) dt
    \Biggr)
\end{aligned}
\end{equation}

where the (column) vector $\bm{\lambda}_{\{j\}}(t)\in\mathbb{R}^{7}$ collects the adjoint variables for each environment. The cost function \eqref{eq:J} and the augmented cost function \eqref{eq:L} are equivalent because the additional term is identically null for any finite choice of $\bm{\lambda}_{\{j\}}$ (by definition of the underlying dynamics in \ref{eq:dxdt}). Therefore, it is possible to make the gradient computation of both cost functions independent of the sensitivities if the adjoint variables are taken as the solution of the following terminal value problem \cite{Navon1998}:
\begin{equation}
\label{eq:dlamdt}
\left\{ \begin{aligned} 
  \frac{d\bm{\lambda}_{\{j\}}}{dt}
    &=
    - \left( \frac{\partial \mathcal{L}}{\partial \bm{s}} \right)_{\{j\}}
    - \bm{\lambda}_{\{j\}}(t)^T
    \left( \frac{\partial \bm{f}}{\partial\bm{s}} \right)_{\{j\}}  \\[0.5em]
  \bm{\lambda}_{\{j\}}(T) &= 0
\end{aligned} \right.
\end{equation}

This is a linear system of ODEs that must be integrated backward in time for $\bm{\lambda}_{\{j\}}(t)$. The gradient can then be computed as 
\begin{equation}
\begin{aligned}
\label{adj_grad}
    \nabla_{\bm{w}} \mathcal{J}(\bm{w})
    =
    \sum_{j=1}^{N}
    \Biggl(
    \int_{t_j}^{t_j+T}
    &\left(
    \frac{\partial \mathcal{L}}{\partial \bm{w}}
    \right)_{\{j\}}
    \\+
    &\bm{\lambda}^T_{\{j\}}(t)
    \left(
    \frac{\partial \bm{f}}{\partial \bm{w}}
    \right)_{\{j\}}
    dt   \Biggl)\,.
\end{aligned}
\end{equation}

Thus, in the adjoint-based approach, the gradient of the loss function is evaluated by solving two systems of ODEs for each environment in the ensemble (i.e., one forward in time to obtain $\bm{s}_{\{j\}}(t)$, and one backward in time to obtain $\bm{\lambda}_{\{j\}}(t)$). 

The vector of parameters is iteratively updated throughout the optimization loop as

\begin{equation}
\label{eq:theta_k}
    \bm{w}^{(k+1)} 
    = 
    \bm{w}^{(k)} 
    - 
    \bm{B}^{(k)}
    \nabla_{\bm{w}} \mathcal{J}(\bm{w}^{(k)})\, , 
\end{equation}

\noindent
where the superscript $^{(k)}$ indicates the iteration counter with $k\in [0,1\dots n_k]$, and $\bm{B}$ is the approximation of the Hessian inverse according to \cite{LBFGSB}.

The gradient computation in \eqref{adj_grad} and the updates in \eqref{eq:theta_k} define two-time scales of the assimilation problem. The first scale is the observation time $T$, linked to the rate at which new information is collected. The second is the learning time scale implicitly defined by the number of optimization iterations $n_k$ carried out before an update on the gradient is requested or is available. The first time scale defines the rate at which the cost function changes because of the dynamics of the system and the potential occurrence of unseen scenarios. The second time scale defines the rate at which the optimizer travels along the parameter space before the cost function changes.

The optimal setting of these scales poses a fundamental question on the impact of the observation time $T$ in relation to the observed scenarios and the quality of the gradient computation: one might ask, for example, whether the observation time should be long enough to observe at least two or three pressurizations or sloshing events. We also investigate the impact of the observation time on the assimilation. 


\section{Results and discussion}
\label{sec6}

We split the presentation of results and the discussion into a section dedicated to the performance of a single environment (Section \ref{sec6p1}) and a section dedicated to multiple environments (Section \ref{sec6p2}).

\subsection{Single-environment performance}
\label{sec6p1}

We first describe the investigated scenario in \ref{sec6p1p1}. Section \ref{sec6p1p2} studies the impact of measurement noise in the collected data for different observation times. Section \ref{sec6p1p3} investigates the role of truncating the observation $T$ to shorter intervals. 
Section \ref{sec6p1p4} reports on the impact of mini-batching the assimilation using only a portion of the data. Finally, section \ref{sec6p1p5} briefly discusses the sensitivity of the system parameters to the closure coefficients and hence the well-posedeness of the inverse problem driving the assimilation.

\subsubsection{Test case description}
\label{sec6p1p1}

\begin{figure*}[h]
	\centering
	\includegraphics[width=0.95\linewidth]{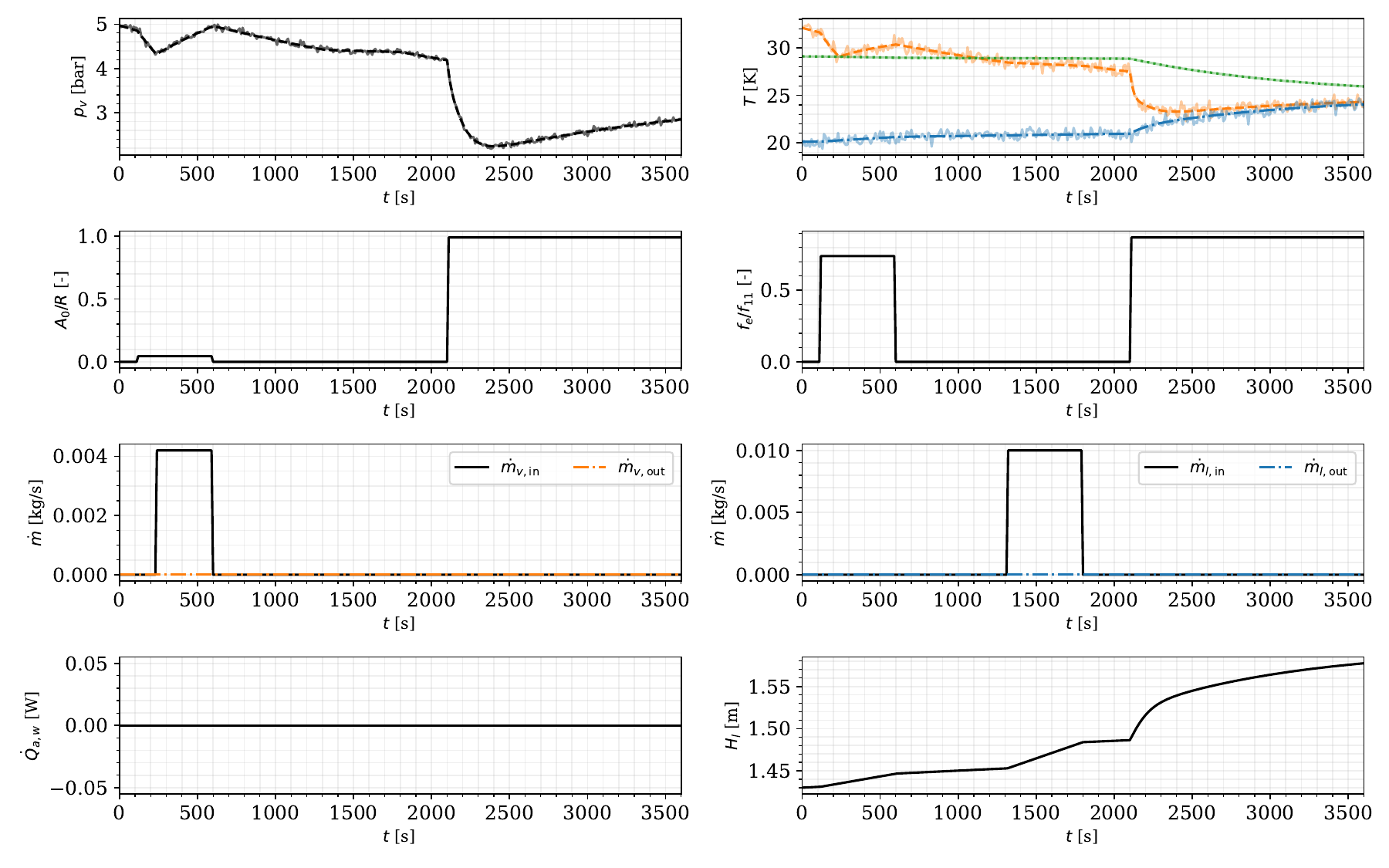}
	\caption{Synthetic data generated for the environment `case 1' used to evaluate the single-environment performance. The components of the observation vector are displayed alongside the injected/removed mass fluxes, the non-dimensional excitation conditions, and the external heat flux.}
	\label{fig:sample_evolution}
\end{figure*}

We consider an upright cylindrical tank with an internal diameter of 1.8 m and a height of 2.2 m. It is filled up to 65\% with LH$_2$ starting from a superheated vapor state at 4.8 bar and 32 K. The liquid is subcooled to 20.09 K, while the tank's material has a heat capacity of $m_w c_w= 2.44$ MJ/K and is initially at 29.1 K. 

Figure \ref{fig:sample_evolution} illustrates the loading and operation scenario of the tank in an observation of 1 hour. In what follows, this denoted as `case 1'. The tank conditions are randomly generated to cover the four scenarios described in Section \ref{sec4}. The first row of plots illustrates the time evolution of the ullage gas pressure (on the left), and the mass averaged temperatures of the vapor, the liquid and the solid volumes (on the right). The second row plots the time amplitude (left) and the frequency (right) of sloshing events. The third row plots the mass inflow/outflow of vapor (left) and liquid (right) as a function of time. Finally, the last row shows the heat flux exchanged through the walls (left) and the liquid level (right). 

A detailed overview of these graphs helps understand the complexity of the system control problem.
In the presented scenario, the tank is undisturbed in the first 2 minutes. Then, a moderate sloshing event occurs with a dimensionless amplitude of $A_0/R=0.045$ and dimensionless frequency $f_e/f_{11}=0.8$ (see definitions of sloshing conditions in Section \ref{sec5p1}). This triggers a visible pressure drop. To counter-balance this, at time $t=4$ minutes, hydrogen vapor at 40 K and 2 bar is injected in the ullage at a rate of 4.2 g/s for 6 minutes. This allows the tank to recover 4.8 bar pressure at 10 minutes when the sloshing event ends. From $t=10$ to $t=22$ minutes, the tank is again undisturbed but the temperature difference between the wall, the gas, and the liquid results in the warming of the liquid and the cooling of the vapor. This slightly reduces the pressure in the ullage. Finally, between $t=22$ and $t=30$ minutes, liquid at 22 K is injected into the system at a rate of 0.01 kg/s. This reduces the ullage volume and produce a moderate compression that arrests the decreasing trend. Finally, at $t=35$ minutes, a violent sloshing event occurs and continues until the end of the observation. This results in a sudden pressure drop due to the significant condensation (see liquid level evolution), followed by a moderate pressure rise due to heat exchanges with the (now) warmer walls. Throughout this test, no heat exchange is assumed to occur from the environment.

\begin{figure*}[!htb]
	\centering
	\includegraphics[width=1\linewidth]{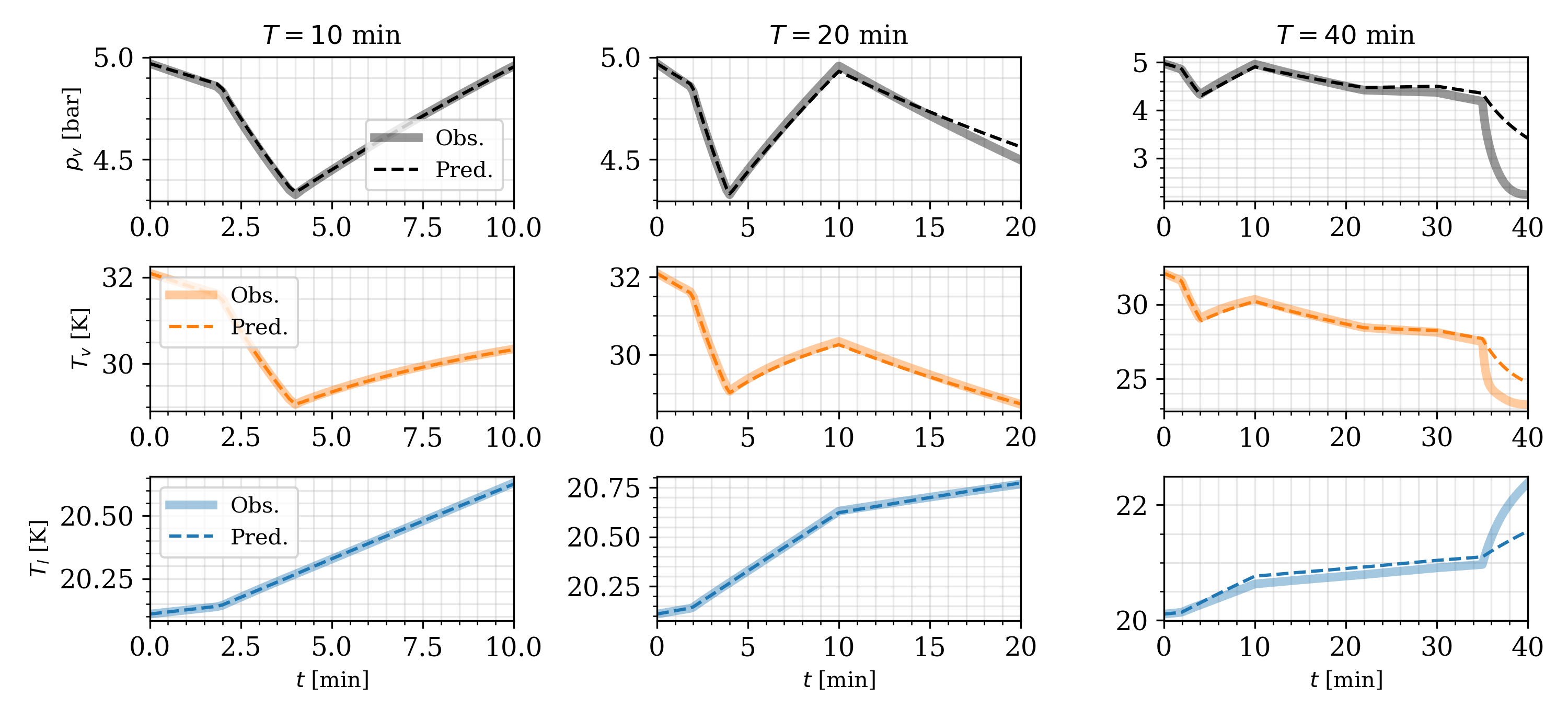}
	\caption{Model predictions and observation data extracted from the test described in Section \ref{sec6p1} for 10, 20 and 40 minutes of observation.}
	\label{fig:1_tank_learning}
\end{figure*}

\begin{figure}[!htb]
	\centering
	\includegraphics[width=1\linewidth]{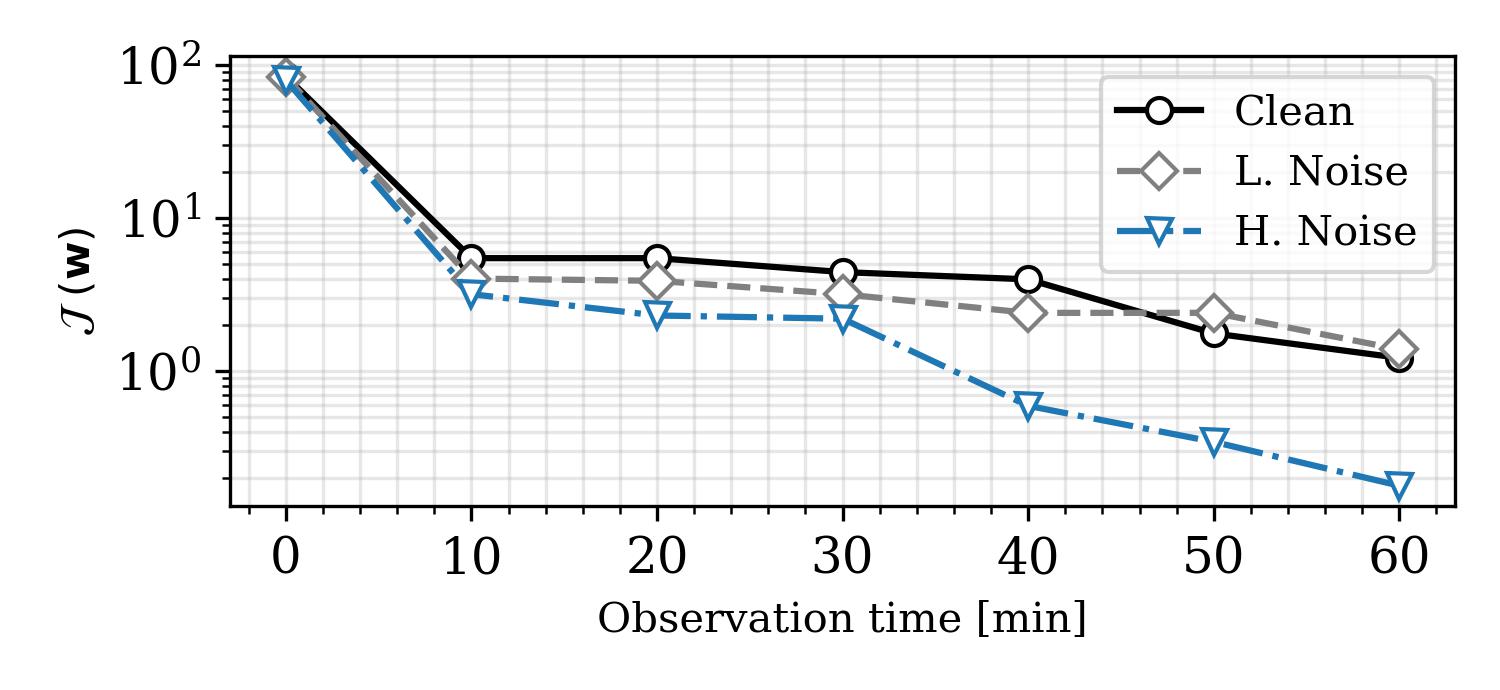}
	\caption{Performance of the data assimilation framework applied to one environment for different noise levels in the data. Minimum of the cost function as a function of the observation time.}
	\label{fig:1_tank_noise}
\end{figure}

\subsubsection{Impact of noise in the observation data}
\label{sec6p1p2}

We here consider three scenarios in terms of measurement noise in the collected observations; these are denoted (1) `clean' (2) `low-noise', and (3) `high-noise'. No noise is present in the first, while Gaussian noise (with zero average) is added in the other two. In (2), the noise's standard deviation is 2 kPa on the pressure, 0.2 K on the temperature, and 0.5 mm on the fill level. In (3), these values are doubled. We considered various observation windows $T$, with increments of 10 minutes, of the environment described in the previous section for the three scenarios. 

The noise impact analysis is completed in Figure \ref{fig:1_tank_noise}, which presents the minimum of $\mathcal{J}$ as a function of the observation time for the three tested scenarios as a function of the observation time. When noise is added, the minimal cost function is expected to be larger in the presence of noise since the underlying model filters it out from the data. Nevertheless, these results show that the optimization is robust to noise, and the exogenous perturbations even led to a slight improvement in the convergence, leading to lower values of $\mathcal{J}$. Our results suggest that the measurement noise helps the optimization avoid local minima, and the parameters identified in the assimilation are learned with shorter observation times.

\begin{figure*}[!htb]
	\centering
	\includegraphics[width=1\linewidth]{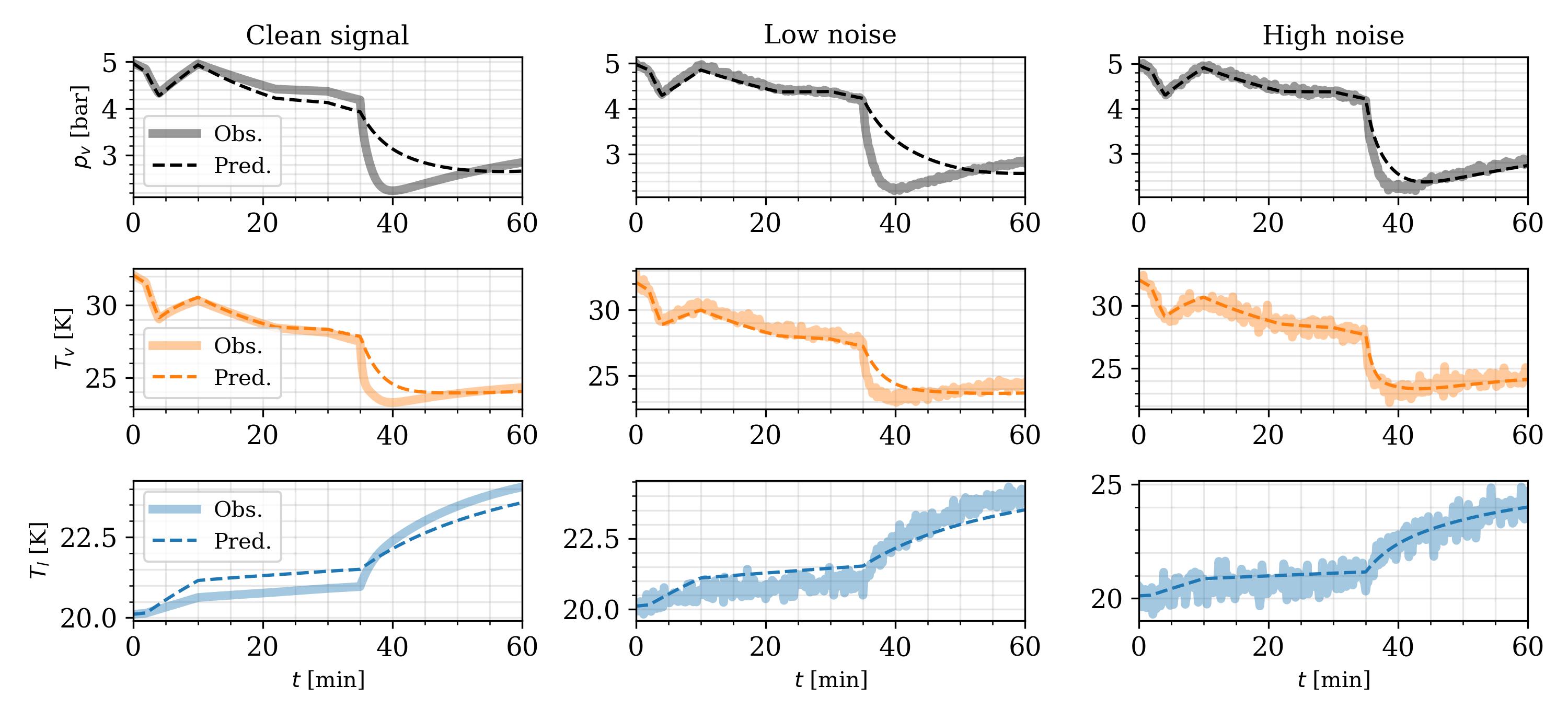}
	\caption{Model predictions and observation data subject to different noise levels for gradually increasing observation windows. The predictions are evaluated at the end of the training set once the full system has been observed for one hour.}
	\label{fig:1_tank_learning_noise_all}
\end{figure*}

Figure \ref{fig:1_tank_learning} shows the results of the real-time data assimilation applied to the `clean' data for $T=10$, $20$, and $30$ minutes. The thick solid lines indicate observations from the virtual experiment, whereas the thin dashed lines indicate the predictions given by data assimilation. The columns of the figure show the tank pressure, vapor temperature, and liquid temperature as the observation time increases.

Interestingly, the assimilation produces excellent matching between the model and observation on short time windows. However, although the effects of mild sloshing, self-pressurization, and fluid injection are well captured, the model fails at accurately predicting the steep thermodynamic variations occurring at $t=35$ minutes due to the violent sloshing event. The same conclusions are drawn on the case with moderate noise as depicted in Figure \ref{fig:1_tank_learning_noise_all} which shows the optimal predictions achieved for the three test cases with the longest observation $T=60$ min. This effect can be interpreted as an over-fitting on the first half of the observations, which is constantly passed to the assimilation. On the other hand, the `high-noise' case shows a better match with the observation. This effect is attributed to the additional stochasticity introduced by the perturbations, helping the optimizer explore a wider parameter space.
 These results also highlight the importance of allowing the agent to witness all possible events: the model/agent cannot predict the consequences of sloshing if this has never been observed in the training data.

\subsubsection{Impact of truncated observation data}
\label{sec6p1p3}

\begin{figure*}[h]
	\centering
	\includegraphics[width=1\linewidth]{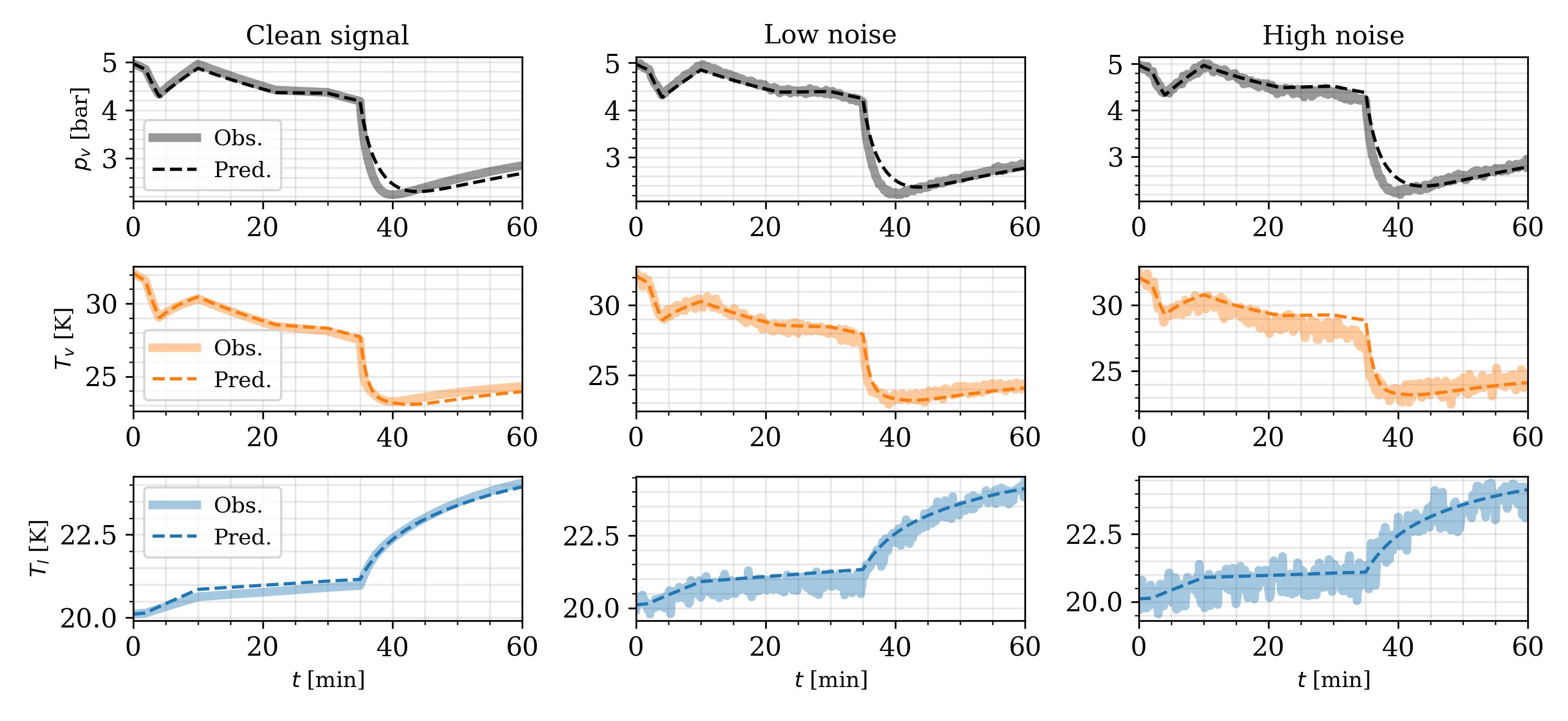}
	\caption{Model predictions and observation data subject to different noise levels for fixed observation windows. The predictions are evaluated at the end of the training set once the full system has been observed for one hour.}
	\label{fig:1_tank_learning_noise_all_short}
\end{figure*}

In this section, we train the agent using only newly observed data with a fixed time window rather than a varying one, including all available observations. This approach mitigates the risk of over-fitting to the initial portion of the observation, as encountered in the previous section. This strategy comes at the risk of reducing the accuracy of the signal portion not included in the assimilation loop. However, as one can see in Figure \ref{fig:1_tank_learning_noise_all_short}, the experience gathered prior to the current observation window is retained by the agent's ANN weights. This can be explained by the significant number of parameters and the relatively low parameter space explored by the optimizer. It is worth noting that this effect was not observed in an earlier version of this work, where only 6 parameters were subject to assimilation \cite{marques_real_2023}.

\begin{figure}[H]
	\centering
	\includegraphics[width=1\linewidth]{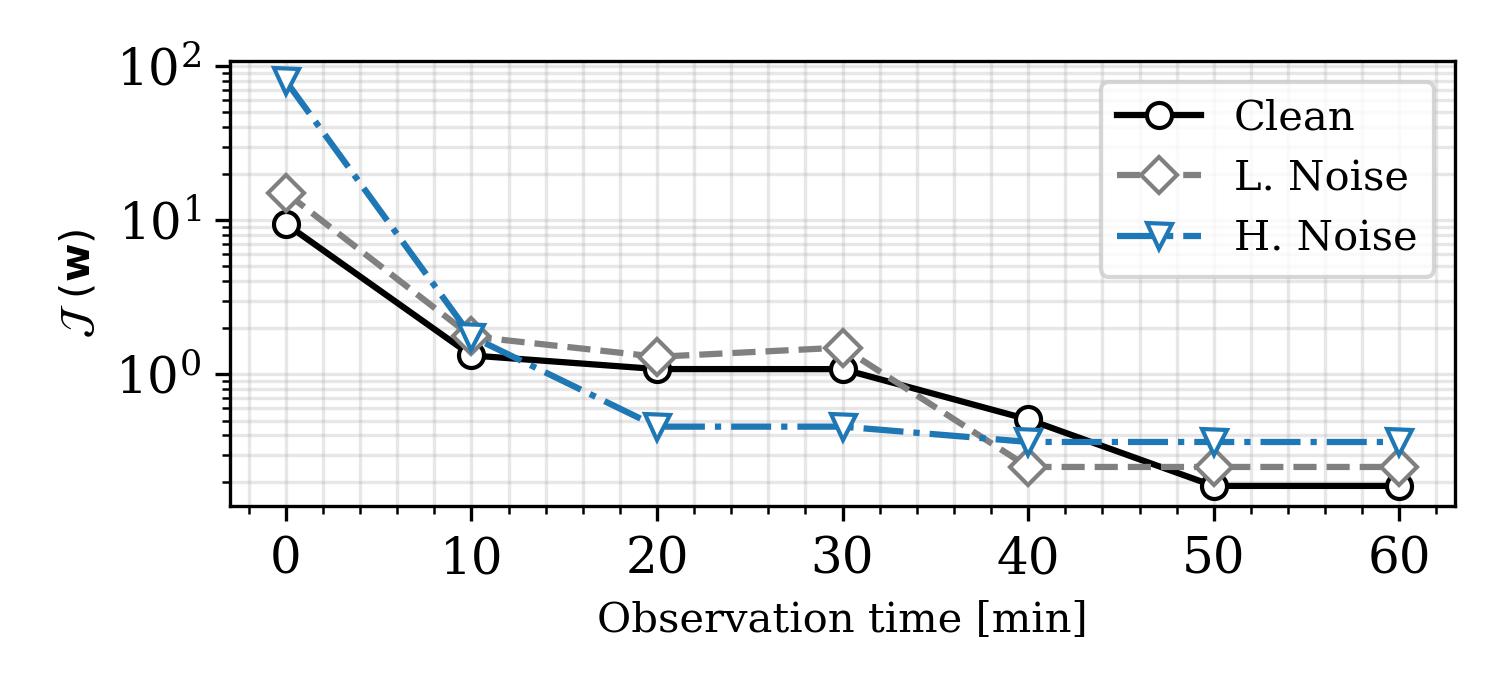}
	\caption{Performance of the data assimilation framework applied to one environment for different noise levels in the data for fixed observation window length, minimum of the cost function as a function of the observation time. }
	\label{fig:1_tank_noise_short}
\end{figure}

The assimilation framework confirms its robustness to noise. However, as depicted in Figure \ref{fig:1_tank_noise_short}, the learning curve saturates to a higher value of loss as the noise increases. This effect is expected as the prediction of the agent remains comprised within the noise uncertainty, while the loss value presented in Figure \ref{fig:1_tank_noise_short} is computed, taking the clean observation as reference. Interestingly, the `clean' and `low-noise' observations are saturating as off 40 minutes, once all the types of external excitations have been witnessed. On the other hand, the `high-noise' case starts saturating at 20 minutes, as only moderate sloshing, vapor injection, and relaxation have been observed.  

\subsubsection{Impact of mini-batch size}\label{sec6p1p4}

The mini-batch sampling is a classic approach to escape local minima and limit memory requirements in the gradient descent method \cite{minibatch}. The idea consists of computing the cost function gradient using a randomly chosen subset of the observation data; this produces an approximation of the gradient $\nabla_{\bm{w}} \mathcal{J}$ that might not always point towards the local minima. 

\begin{figure}[h]
	\centering
	\includegraphics[width=1\linewidth]{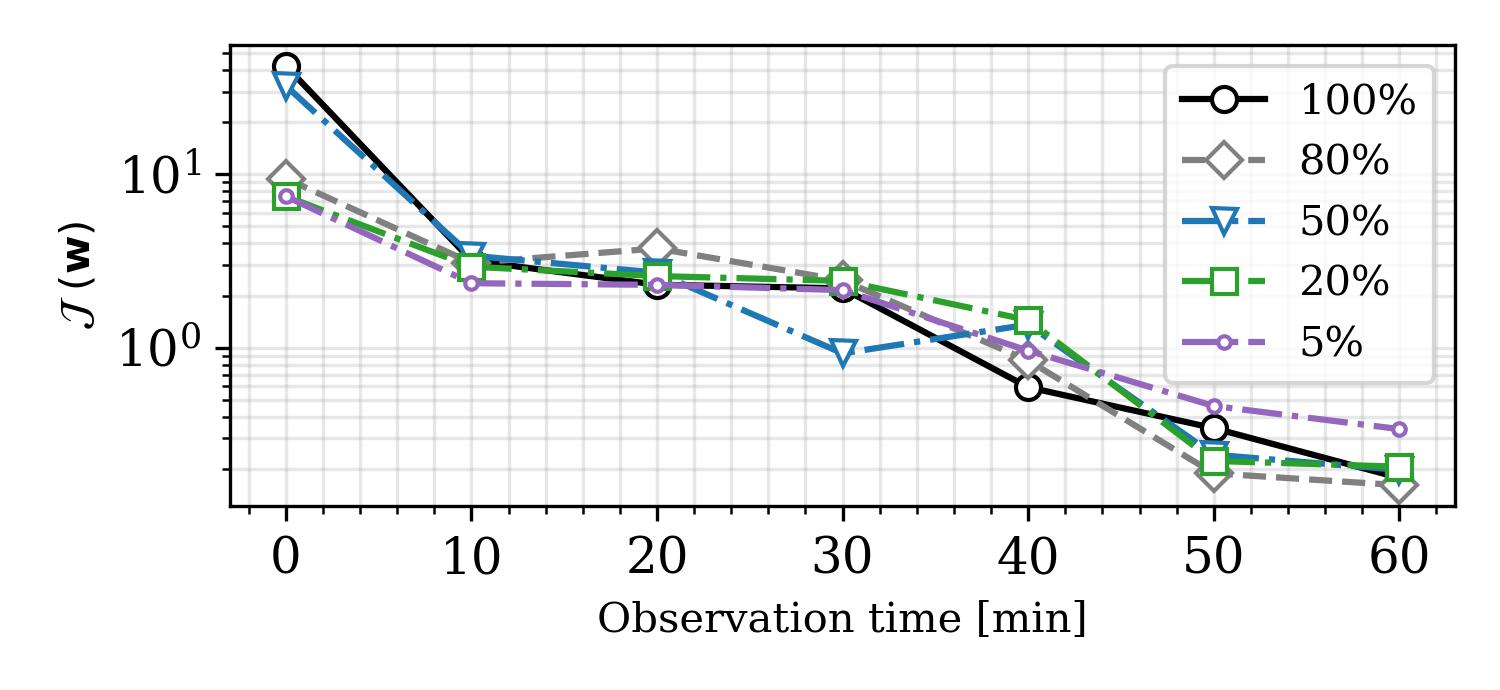}
	\caption{Performance of the data assimilation framework applied to one environment, minimum of the cost function as a function of the observation time for different size of mini-batch. }
	\label{fig:1_tank_minibatch}
\end{figure}

The use of mini-batch strategies on quasi-Newton methods, such as the L-BFGS-B technique used in this work, is the subject of active research (see \cite{Bollapragada2018}). Nevertheless, in this work, we explored its impact on the `high-noise' configuration from the previous subsection. We test the assimilation using 80\%, 50\%, 20\% and 5\% of the observed data to evaluate $\nabla_{\bm{w}}\mathcal{J}$. To account for the stochasticity of the process, the assimilation is repeated one hundred times in each case, and the results are averaged. The average behavior of the cost function evolution is shown in Figure \ref{fig:1_tank_minibatch}. Remarkably, the additional stochasticity of the mini-batch selection slightly improves the result of the assimilation. This approach appears to provide the best performance as 80\% of the observation is sampled. On the other hand, an extreme down-sampling of the observations (5\%) has a slight negative impact on the assimilation results.

In summary, these results showcased that sampling portions of the training data during the optimization can improve the assimilation results. Furthermore, this sampling provides an approximation of $\nabla_{\bm{w}}\mathcal{J}$, which allows for a broader exploration of the parameter space, yielding better predictions for $\bm{w}$, even if the value of the cost function is not significantly affected (on average). For the current training data set, the best compromise was obtained by sampling 80\% of the input data.

\subsubsection{Well-posedness and sensitivities}
\label{sec6p1p5}

In this section, we assess the sensitivity of model predictions $\bm{y}(t)$ to the closure parameters $\bm{\theta}$ (heat transfer coefficients). We use a standard Sobolev indices analysis
\cite{Sobol2001}. This consists in treating the terms in $\bm{\theta}$ as random variables and analyzing how their distribution propagates through the thermodynamic model.

In practice, for each of the parameters in $\bm{\theta}$, we construct a Gaussian distribution with mean centered on the expected value (ground truth for the data generation) and standard deviation equal to 10\% of the mean. Keeping the other parameters equal to the mean, we randomly sample 100 possible outcome of a given coefficient and compute the associated thermodynamic evolution from the model in section \ref{sec2}.

\begin{figure}[h]
	\centering
	\includegraphics[width=1\linewidth]{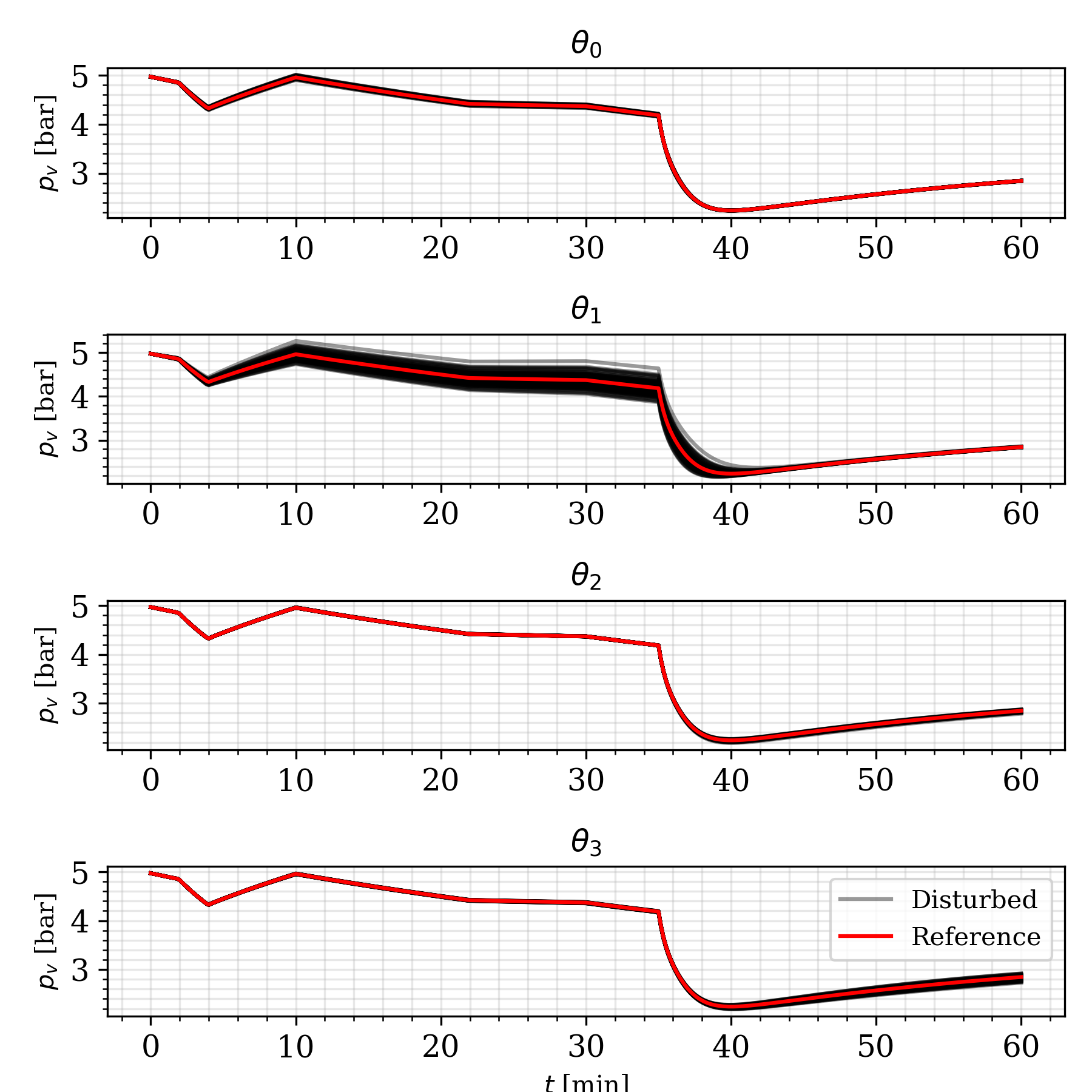}
	\caption{Sensitivity of the pressure prediction on the vector of model parameters $\bm{\theta}$ with a 10\% variation for `case 1'. From top to bottom, $\theta_0$ to $\theta_3$ }
	\label{fig:h_coef_sensitivity}
\end{figure}

Figure \ref{fig:h_coef_sensitivity} shows the predicted pressure obtained through this investigation, applied to `case 1'. The red lines correspond to the prediction obtained with the baseline coefficients (i.e., sampling the mean value of each distribution). The black lines correspond to the random samples of $\bm{\theta}$. The rows of the figure correspond to the random sampling for (1) $\theta_0=h_{v,i}$, (2) $\theta_1=h_{l,i}$, (3) $\theta_2=h_{w,v}$, and (4) $\theta_3=h_{w,l}$, respectively. 

The results show that predicted pressure is predominantly affected by variations in $\theta_1$ (i.e., the heat transfer coefficient between the liquid and the interface $h_{l,i}$). 
This is explained by the fact that the liquid governs the system thermodynamic when mixing occurs, as it encloses more energy than the vapor phase. The role of this coefficient is particularly important in the relaxation phase and the case of sloshing.

During the first relaxation phase ($t = 10$ min to $t = 22$ min), in fact, the temperature difference between the gas and the walls is negligible (see Figure \ref{fig:sample_evolution}) and the heat exchange happens mostly at the interface. This explains also the negligible impact of $h_{w,v}$ and $h_{w,l}$ during these phases.

The situation is inverted after the violent sloshing event ($t > 35$ min). The thermal mixing between vapor and liquid brings these to nearly the same temperature while the walls have a slower reaction time and remain superheated. In this phase, as the exchanges with the wall are more important than those at the interface, the sensitivity to parameters $\theta_0$ and $\theta_1$ decreases in favor of $\theta_2$ and $\theta_3$, with the latter being more important because of the larger heat capacity of the liquid with respect to the vapor.

The model sensitivities to these parameters define the well-posedness of the inverse method underlying the assimilation and link the various scenarios to possible learning opportunities for the agent. For example, the violent sloshing event in this test case would be a learning opportunity for $\theta_1$ but not $\theta_2$ because the thermal evolution of the tank during this phase depends strongly on the first and poorly on the second. The optimization driving the assimilation has thus a stronger gradient on the first and a nearly vanishing gradient on the second. 

From a more formal point of view, the problem of identifying $\theta_2$ in a continuously sloshing tank is thus not well posed because the lack of sensitivity implies that a large range of $\theta_2$ could be associated with the same observed pressure evolution. This is why the multi-environment formalism proposed in this work is particularly attractive.


\begin{figure}[h]
	\centering
	\includegraphics[width=1\linewidth]{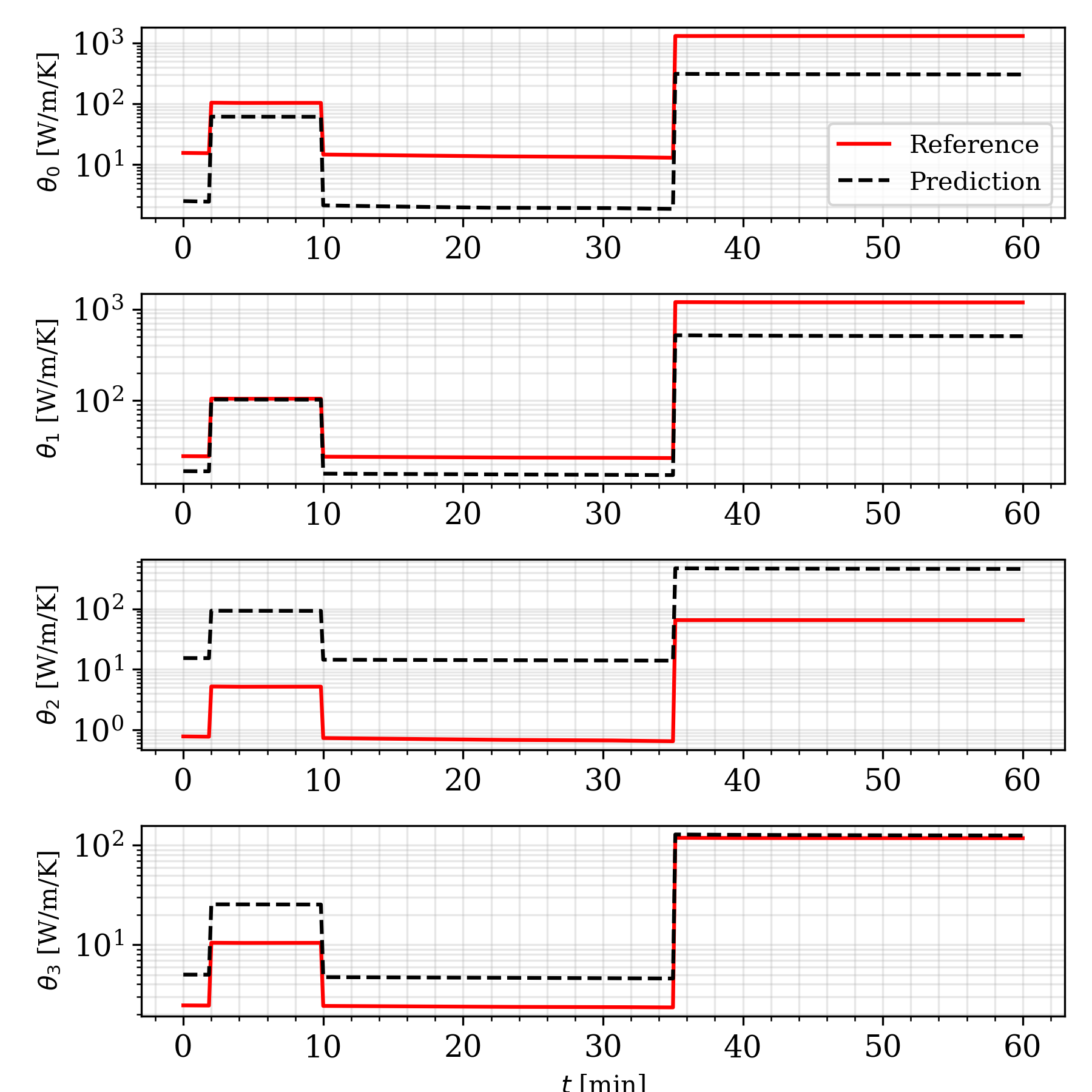}
	\caption{Time evolution of the reference and predicted heat transfer coefficients, for an agent with high noise level and fixed observation windows. From top to bottom, $\theta_0$ to $\theta_3$. The }
	\label{fig:h_coef}
\end{figure}

Finally, to further highlight the well-posedness problem, Figure \ref{fig:h_coef} shows the predicted heat transfer coefficients for `case 1' using `high noise' and fixed observation windows. The red continuous lines show the reference (expected) values while the black dashed lines show the predicted ones.

The results illustrate the impact of the model sensitivity on the identification of the heat transfer coefficients: while an excellent match on the observation quantities (see Figure \ref{fig:1_tank_learning_noise_all_short}) was retrieved, a significant mismatch on the predicted parameters $\bm{\theta}$ is observed in those time interval where the model sensitivity is too low. Therefore, the parameter $\theta_1=h_{l,i}$ is remarkably well identified in the time interval from $t=2$ min to $t = 10$ min, and the same is true for the parameter $\theta_3=h_{w,l}$ from $t>35$ minutes. The identification performances on the others are poorer, with the worst results obtained for $\theta_0=h_{v,i}$ which has the weakest impact on the model evolution.

While the proposed approach seeks to tackle these limitations with a multi-environment formalism, in the hope that these sufficiently span all learning opportunities for the assimilation agent, future developments could include a gradient re-scaling approach to account for the model sensitivity.

\subsection{Multi-environment performance}
\label{sec6p2}

We consider the assimilation from multiple tanks (environments) with different sloshing profiles, thermal loading, and inflow/outflow control actions. These are generated with random sequences of scenarios like the `case 1' described in Section \ref{sec6p1p1}.

\begin{figure*}[h!]
	\centering
	\includegraphics[width=0.95\linewidth]{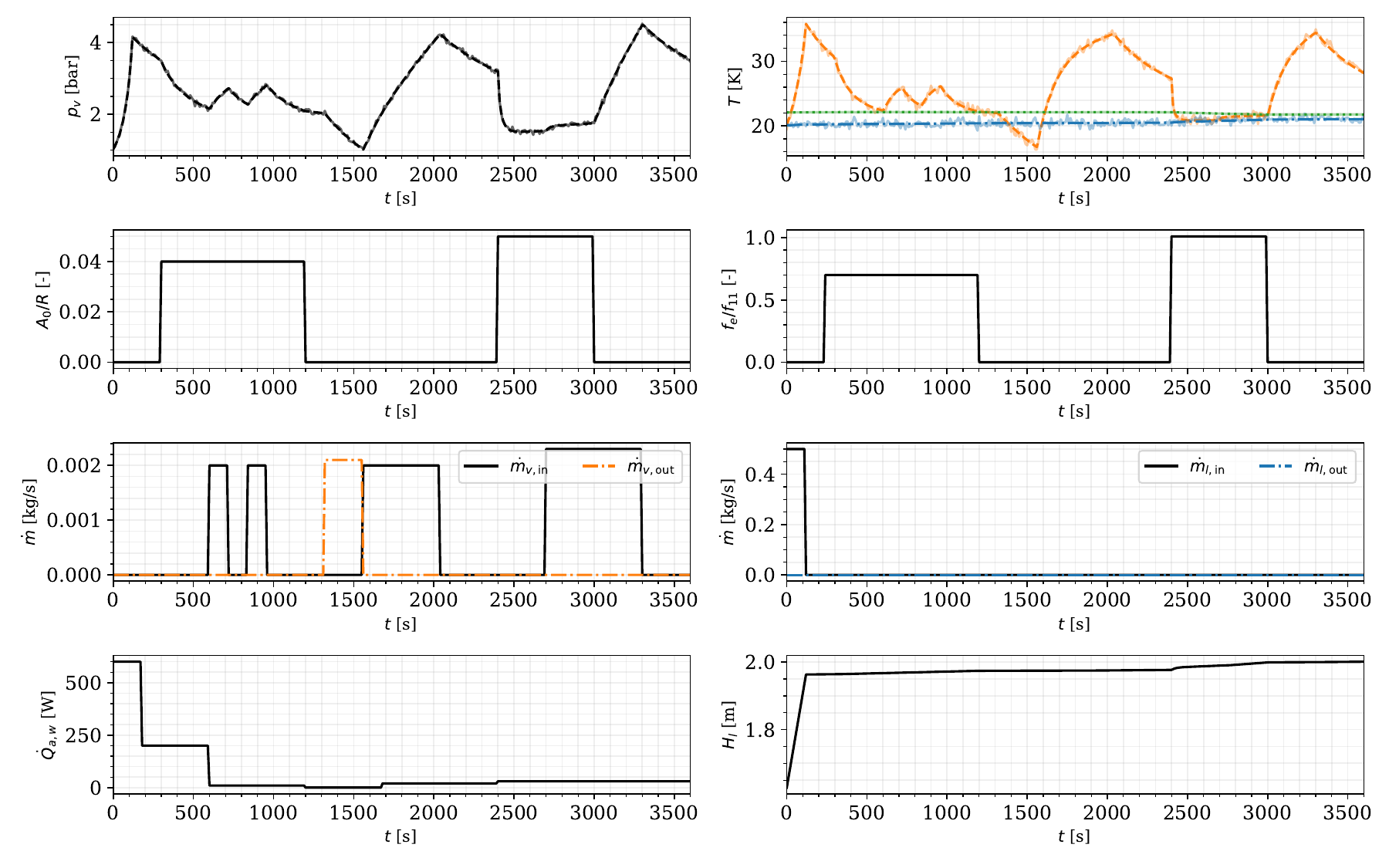}
	\caption{Thermodynamic evolution and operating sequence of case 2.}
	\label{fig:app_tank_2}
\end{figure*}

\begin{figure*}[h!]
	\centering
	\includegraphics[width=0.95\linewidth]{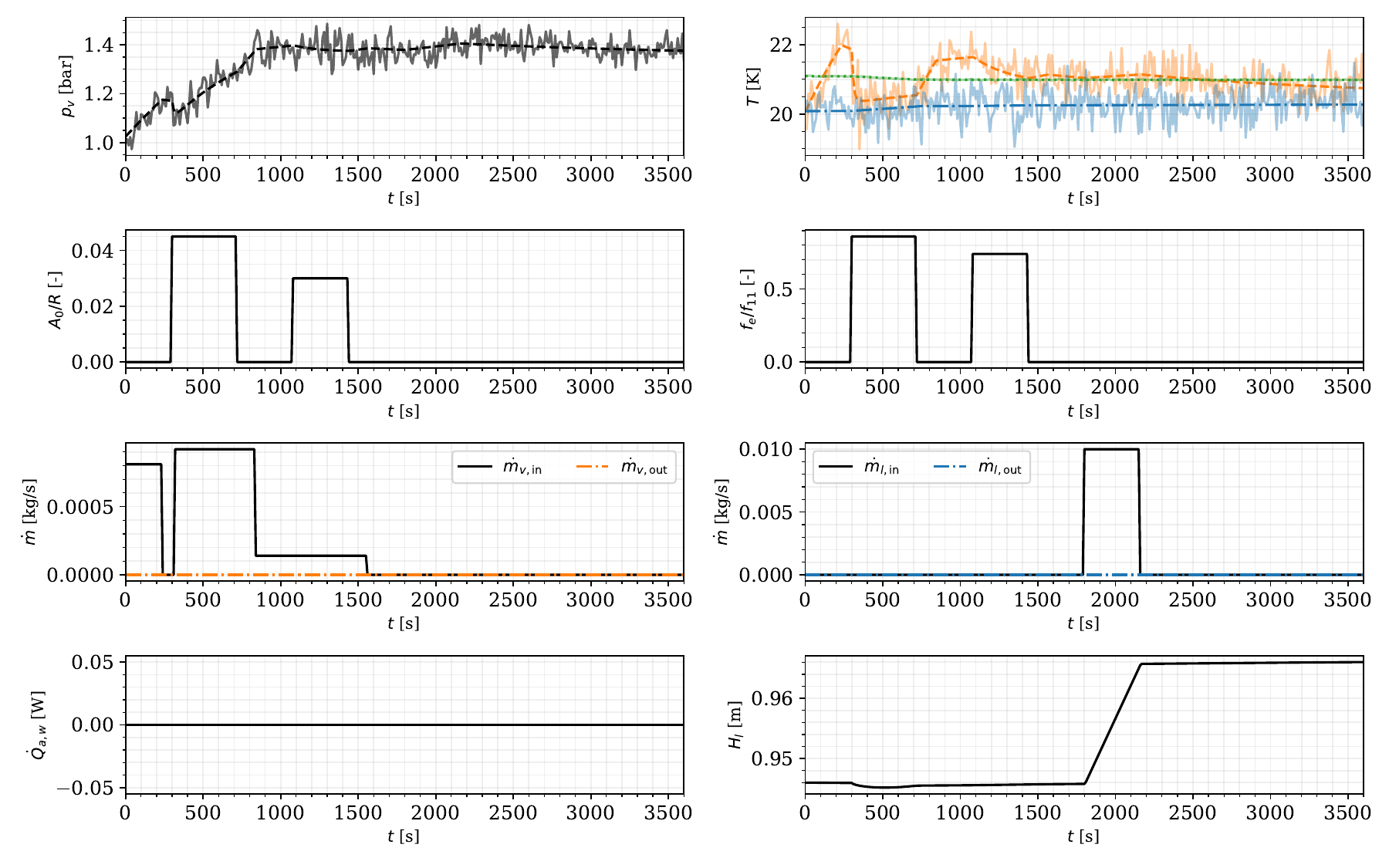}
	\caption{Thermodynamic evolution and operating sequence of case 3.}
	\label{fig:app_tank_3}
\end{figure*}

\begin{figure*}[h!]
	\centering
	\includegraphics[width=0.95\linewidth]{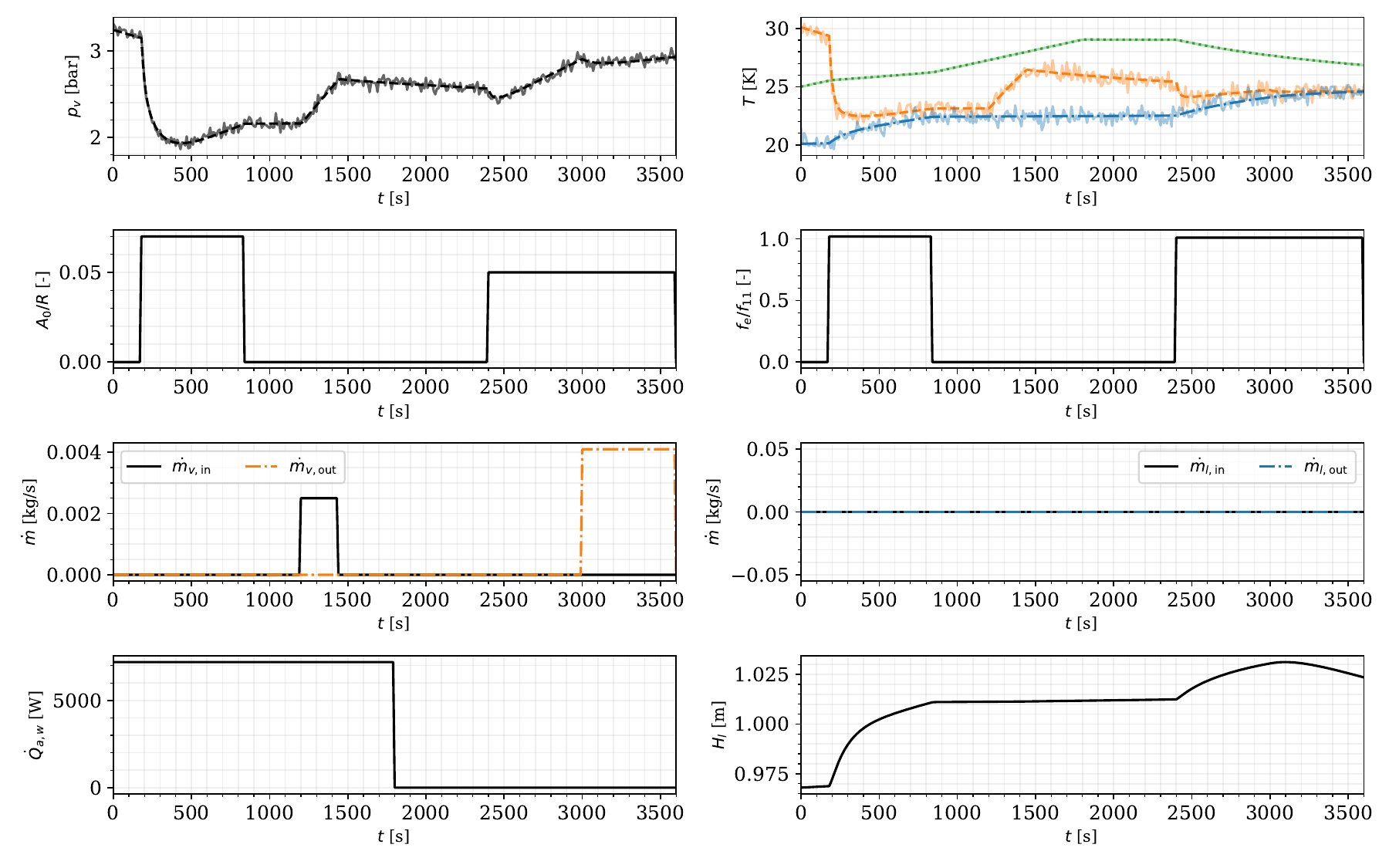}
	\caption{Thermodynamic evolution and operating sequence of case 4.}
	\label{fig:app_tank_4}
\end{figure*}

\begin{figure*}[h!]
	\centering
	\includegraphics[width=0.95\linewidth]{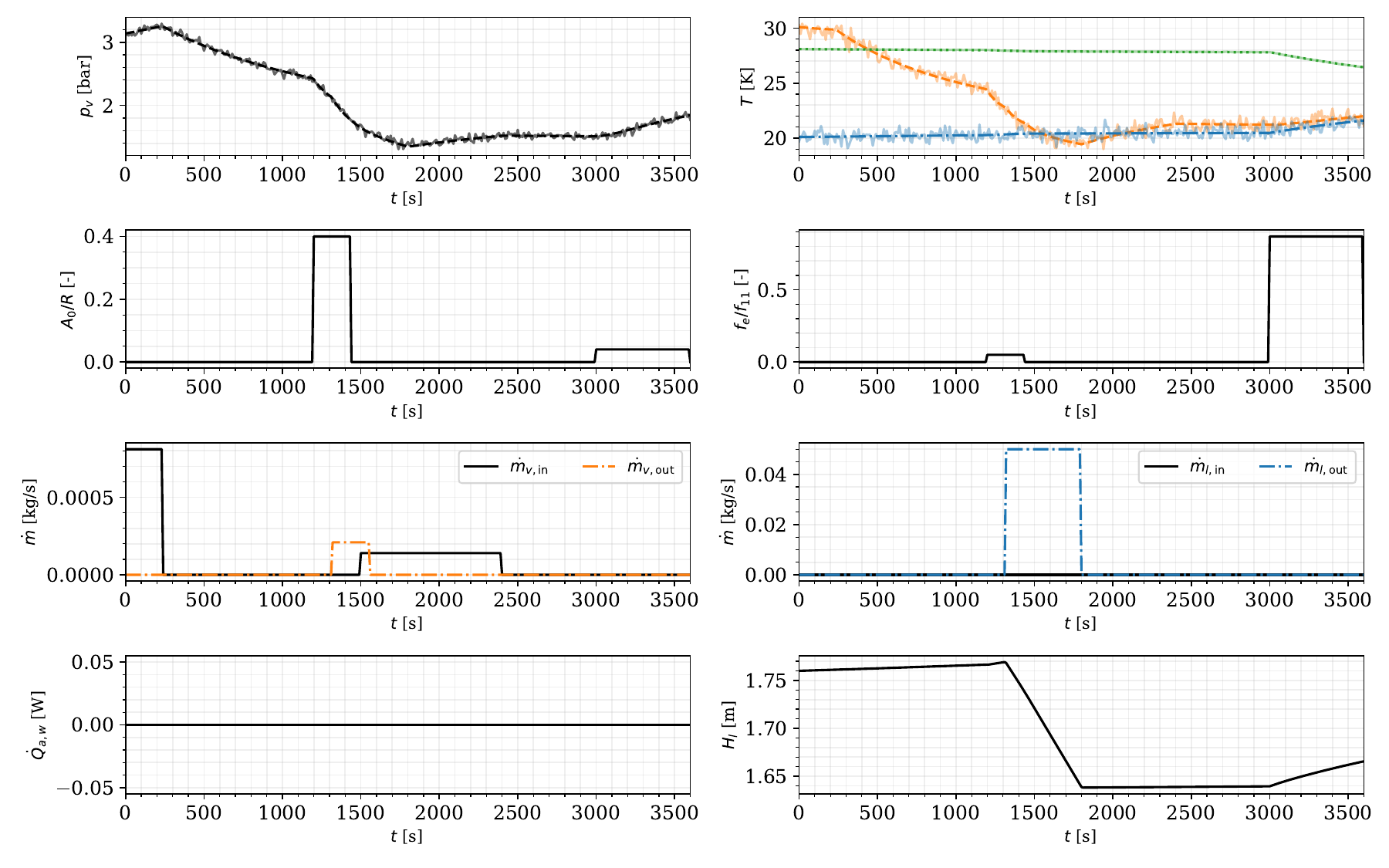}
	\caption{Thermodynamic evolution and operating sequence of case 5 (i.e., the validation tank).}
	\label{fig:app_tank_5}
\end{figure*}

The operation of these new cases is summarized in Figures \ref{fig:app_tank_2}-\ref{fig:app_tank_5} in a similar fashion as Figure \ref{fig:sample_evolution}. All environments are observed for the same total duration of $T=60$min. Like `case 1' in Section \ref{sec6p1p1}, all cases consider LH$_2$ initially at 20.09 K and have the same geometry. However, these undergo radically different sequences of operations. Case 2 starts with the ullage in near saturation condition ($T_v = 20.1$ K) and slightly superheated walls ($T_w=22.09$ K), with a fill ratio of 78\%. The tank initially receives a strong heat ingress, which goes from 600W to about 10 W. From $t = 0$ to $t = 2$ min, the fill level is increased by liquid injection, leading to a rise of the tank pressure. From this point onward, the tank experienced two violent sloshing events (the first between 5-20 min, and the second between 40-50 min), four vapor injections, and one venting operation. Among all cases, this test case is the one featuring the largest sloshing-induced pressure fluctuations.

`Case 3' also started near saturation conditions with the same initial vapor temperature as `case 2', albeit with a slightly lower wall temperature of $T_w = 21.09$ K, and a much lower fill-ratio of 43\%. In contrast to the previous case, no external heat flux was considered here, and sloshing events are significantly milder. This case was also characterized by slow vapor injections which lasted until $t = 26$ min. This is the test case that experienced the least pressure fluctuations. 

Finally, `case 4' starts in super-heated conditions with $T_v = 30.1$ K, $T_w = 25$ K, and a fill-ratio of 44\%. This environment faces a large external heat flux of 7.2 kW until $t=30$ min. This test case also experiences two violent sloshing events. The first produces a strong pressure drop, while the second is counter-acted by injecting vapor in the ullage, leading only to a mild pressure fluctuation.

In addition to these cases, we include a 5th case, which is not used for training the agent but only for validation purposes. 
This is characterized by initially superheated conditions in the vapor, initial wall temperatures at $T_w=28.1$, and a fill-ratio of 80 \%. The tank is initially pressurized until a $p_v=3.1$ bar, then put on hold, undisturbed until $t=20$ min, a high amplitude and low frequency sloshing disturbance is applied. This reduces the pressure and promotes mixing between the warmer vapor and the colder liquid until they collapse to nearly the same temperature. This pressure drop is counterbalanced by a hot vapor injection from $t=25$ to $t=40$ min. Following this, some liquid is removed between $t=22$ and $t=30$ min to prevent the liquid level from rising too much due to condensation. The test case finishes with a high-frequency excitation from $t>50$ min. Consequently, the homogenized gas and liquid are gradually warmed up by the superheated walls, producing a gradual pressure rise.

\begin{figure}[h]
	\centering
	\includegraphics[width=1.0\linewidth]{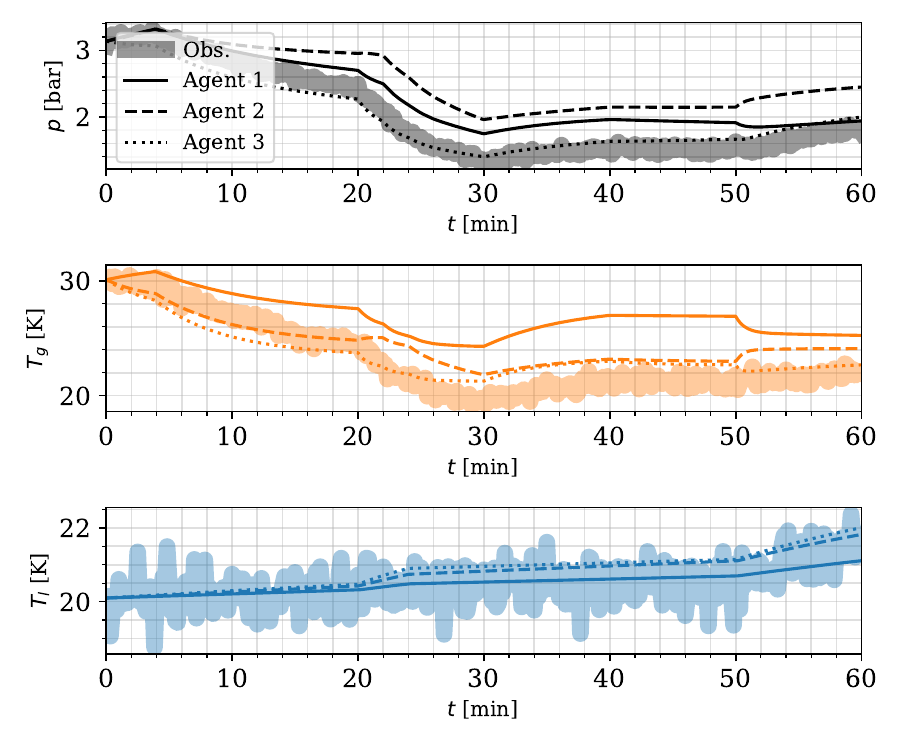}
	\caption{Pressure predictions and observation data for different number of tanks used in the training of the agent}
	\label{fig:pressure_multitank}
\end{figure}

To analyze the impact of the multi-environment framework we compare the performances of three agents, trained with different environments. Agent 1 is trained using only $N=1$ environments (`case 1'), Agent 2 is trained using $N=2$ environments (`cases 2 and 3'), and Agent 3 is trained with $N=3$ environments (`cases 2, 3 and 4'). In all tests, the weights of the agent are initialized using the best-performing ones obtained from the single-tank optimization phase (i.e., `case 1'). We carried out the training for all cases by applying `high noise' to the observations, with batches of 80 \%, and using fixed observation windows of $T=10$ minutes. 

The performances of the three agents are evaluated on the validation test case. Table \ref{tab:loss_multiple_tanks} collects the value of the cost function for the three agents while Figure \ref{fig:pressure_multitank} shows the prediction performance of the three agents on the validation test case. Agent 3, who has learned from multiple environments clearly performs better.



\begin{table}[h]
\centering
\renewcommand{\arraystretch}{1.5}
\caption{Value of the loss function $\mathcal{J}(\bm{w})$ on the validation set for different number of tanks used in assimilation.}
\begin{tabular}{ccc}
\\
\hline
Agent 1  & Agent 2   & Agent 3  \\ \hline
 6.46   &  4.38 &   0.90     \\
 \hline
\end{tabular}
\label{tab:loss_multiple_tanks}
\end{table}



\section{Conclusions}\label{sec7}

We presented a multi-environment real-time data assimilation framework for the data-driven thermodynamic modeling of cryogenic fuel tanks. 

The implemented approach sets up an optimization problem in which the error between system observations and model predictions must be minimized by acting on closure parameters. The model is based on a lumped formulation of mass and energy balances for each of the system's control volumes (i.e., the vapor, liquid, and solid walls), and the optimization is carried out by combining a quasi-Newton approach with the adjoint-based evaluation of the cost function gradients.

The training data is collected by multiple environments, i.e. different tanks undergoing various operating conditions (e.g., sloshing, filling, etc.). To provide the first proof of concept of the proposed framework, the environments are simulated using synthetic empirical correlations. The assimilation aims to train an Artificial Neural Network (ANN) with no embedded knowledge of the sloshing dynamics to identify a closure law from the state of a 0D thermodynamic model to the heat transfer coefficient, using noisy pressure and temperature measurements collected in real-time from multiple tanks.

Various tests were performed for the data assimilation applied to a single environment: we assessed the impact of measurement noise on the training data and the effect of `mini-batches' in the gradient computation alongside the duration of an observation within which the assimilation is carried out. Remarkably, it was found that measurement noise improves the data assimilation, allowing for better identification of the underlying coefficients in shorter observation times. In addition, reducing the observation window to the newly observed data produced systematic improvements in the assimilation convergence while reducing the computational cost of the assimilation procedure. Similarly, the mini-batch sampling approach brought an improvement to the convergence with an optimum found at 80\% of the observation length. Finally, further improvements in the assimilation were achieved in a multi-environment scenario considering three or four tanks undergoing widely different loads and control scenarios.

To conclude, this work provides a solid proof of concept for applying a multi-environment data assimilation framework for the data-driven calibration of thermodynamic models of cryogenic tanks, potentially enabling model predictive control in thermal management systems. An agent with no prior information on heat and mass transfer correlations for the non-isothermal sloshing proved able to "learn" the relevant laws and predict, with the help of a thermodynamic model, the evolution of relevant parameters in a storage tank. In the era of the fourth industrial revolution, where intelligent sensors are becoming widely available, the perspective of deploying the proposed multi-environment assimilation to a large fleet of heavily instrumented tanks appears particularly promising.

\section*{Acknowledgments}

This work has been funded by the Flemish Agentschap Innoveren \& Ondernemen in the framework of the CSBO project ``Clean Hydrogen Propulsion for Ships (CHyPS)''. Pedro Marques is supported by the FRIA grant 40009348 from the `Fonds de la Recherche Scientifique (F.R.S. -FNRS)'.




\bibliographystyle{abbrv}
\bibliography{Marques_et_al_2023}

\end{document}